\RequirePackage{fix-cm}
\documentclass[twocolumn, epjc3]{svjour3}  
\smartqed  
\RequirePackage{graphicx}
\RequirePackage{mathptmx}      
\RequirePackage{latexsym}
\RequirePackage[numbers,sort&compress]{natbib}
\RequirePackage[colorlinks,citecolor=blue,urlcolor=blue,linkcolor=blue]{hyperref}
\RequirePackage[separate-uncertainty=true]{siunitx}
\RequirePackage{placeins}
\RequirePackage{amsmath}
\RequirePackage{microtype}

\newcommand{\MG}{\textsc{MadGraph5}\_aMC@NLO\ }
\DeclareMathAlphabet{\mathcal}{OMS}{cmsy}{m}{n}
\journalname{Eur. Phys. J. C}
\begin{document}

\title{Constraining top-quark couplings combining top-quark and $\boldsymbol{B}$  decay observables}
\subtitle{\hfill \normalfont \normalsize DO-TH 19/17}
\titlerunning{Constraining top-quark couplings combining top-quark and $B$ decay observables}

\author{Stefan Bi\ss{}mann \thanksref{e1,addr1} \and
        Johannes Erdmann \thanksref{e2,addr1} \and
        Cornelius Grunwald \thanksref{e3,addr1} \and
        Gudrun Hiller  \thanksref{e4,addr1}\and
        Kevin Kr\"oninger \thanksref{e5,addr1}
        }

\thankstext{e1}{e-mail: \href{mailto:stefan.bissmann@tu-dortmund.de}{stefan.bissmann@tu-dortmund.de}}
\thankstext{e2}{e-mail: \href{mailto:johannes.erdmann@tu-dortmund.de}{johannes.erdmann@tu-dortmund.de}}
\thankstext{e3}{e-mail: \href{mailto:cornelius.grunwald@tu-dortmund.de}{cornelius.grunwald@tu-dortmund.de}}
\thankstext{e4}{e-mail: \href{mailto:ghiller@physik.uni-dortmund.de}{ghiller@physik.uni-dortmund.de}}
\thankstext{e5}{e-mail: \href{mailto:kevin.kroeninger@tu-dortmund.de}{kevin.kroeninger@tu-dortmund.de}}


\institute{ Fakultät  Physik, TU Dortmund, Otto-Hahn-Str.4, 44221 Dortmund, Germany \label{addr1}
}

\date{Received: date / Accepted: date}

\maketitle
\begin{abstract}
We present a first, consistent combination of measurements from top-quark and $B$ physics to constrain top-quark properties within the Standard Model Effective Field Theory (SMEFT). We demonstrate the feasibility and benefits of this approach and detail the ingredients required for a proper combination of observables from different energy scales.
Specifically, we employ  measurements of the $t\bar t\gamma$ cross section together with  measurements of the $\bar B\rightarrow X_s\gamma$ branching fraction to test the Standard Model and look for new physics contributions to the couplings of the top quark to the gauge bosons within SMEFT.
We perform fits of three Wilson coefficients of dimension-six operators considering only the individual observables as well as their combination to demonstrate how the complementarity between top-quark and $B$ physics observables allows to resolve ambiguities and significantly improves the constraints on the Wilson coefficients.
No significant deviation from the Standard Model is found with present data.

\end{abstract}

\section{Introduction}
The experiments at the Large Hadron Collider (LHC) conduct various searches for physics beyond the Standard Model (BSM). 
The searches for direct production of new particles have not yet resulted in any discovery of BSM physics. A complementary approach are indirect searches, where precise measurements of total rates and kinematic distributions 
are compared to their Standard Model (SM) predictions. 
If the new particles are heavier than the experimental energy scale, the Standard Model Effective Field Theory (SMEFT) can be applied to parametrize potential deviations from the SM in a model-independent way \cite{Weinberg:1978kz,Buchmuller:1985jz,Grzadkowski:2010es}. For energies below the scale of BSM physics, $\Lambda$, effects of new particles and interactions can be described in a series of higher-dimensional operators constructed from SM fields.
\par
The top quark plays a special role in SMEFT analyses and a large number of precision measurements regarding top-quark physics have been performed at the LHC.
As the top quark is the only fermion with an $\mathcal{O}(1)$ Yukawa coupling, it is of special interest in BSM scenarios explaining the origin of electroweak symmetry breaking (EWSB).
For these reasons, numerous SMEFT analyses in the top-quark sector have been performed during the recent past, for example, \cite{Degrande:2018fog,Chala:2018agk,Durieux:2014xla,AguilarSaavedra:2010zi,DHondt:2018cww,Durieux:2018ggn,Buckley:2015nca,Buckley:2015lku,deBeurs:2018pvs,Brown:2019pzx,AguilarSaavedra:2018nen,Hartland:2019bjb,Maltoni:2019aot,Durieux:2019rbz,Neumann:2019kvk}. 
In particular, first global studies have been presented in Refs. \cite{Buckley:2015nca,Buckley:2015lku,AguilarSaavedra:2018nen,Brown:2019pzx,Hartland:2019bjb,Durieux:2019rbz}. 
\par
Additional constraints on BSM contributions to top-quark physics come from $B$ physics (see e.g. Refs. \cite{Fox:2007in,Grzadkowski:2008mf,Drobnak:2011aa}). Especially flavor-changing neutral currents are excellent probes of BSM physics due to suppression by the Fermi constant, small CKM matrix elements and loop factors. 
The Weak Effective Field Theory (WET) Lagrangian describing $b\rightarrow s$ transitions is not invariant under the full SM gauge group due to EWSB at the scale $v$. Since the scale $\Lambda$ has to be above $v$, BSM physics needs to be integrated out before EWSB. 
To constrain SMEFT coefficients using low-energy observables, the effective Lagrangian must be matched onto the WET Lagrangian by integrating out all particles heavier than the $b$ quark \cite{Aebischer:2015fzz,Fox:2007in,Drobnak:2011aa,Grzadkowski:2008mf,Hurth:2019ula}. 
\par
Matching and renormalization group equation (RGE) evolution enable to combine measurements at different energy scales in one analysis that allows to investigate the impact of measurements from top-quark and $B$ physics on the top-quark sector of SMEFT.
\par
In this paper, we consider $t\bar t\gamma$ cross sections and the $\bar B\rightarrow X_s\gamma$ branching fraction to perform a first consistent fit of SMEFT Wilson coefficients using a combination of top-quark and $B$ physics observables that have a common set of relevant dimension-six operators. Similar analyses have been performed for top-Higgs couplings in Refs.~\cite{Cirigliano:2016njn,Cirigliano:2016nyn}.
We present the steps necessary for such a combined analysis of BSM contributions to top-quark interactions and highlight possible pitfalls in this procedure. 
We determine the dependence of the observables on the Wilson coefficients and compare our computations to results obtained with existing tools. 
We estimate the gain in the sensitivity for BSM contributions when considering top-quark and $B$ physics observables in a combined fit.
\par
The outline of this paper is as follows. In Sec. \ref{Sec:SMEFT} we introduce the SMEFT and WET Lagrangians and introduce conventions used throughout this paper. 
In Sec. \ref{Sec:Match} we discuss the steps necessary to calculate low energy observables in dependence of SMEFT Wilson coefficients.
The measurements used to constrain the SMEFT Wilson coefficients are presented in Sec. \ref{Sec:Measurements}. In Sec. \ref{Sec:Modelling} we describe the corresponding computations of the SM and BSM contributions. In Sec. \ref{Sec:Fit} we determine constraints on the SMEFT Wilson coefficients. We investigate the individual impact of top-quark and $B$ observables and demonstrate how the combination of these observables improves the constraints. 
In Sec. \ref{Sec:Conclusion} we conclude. Auxiliary information is given in several appendices.

\section[Effective field theories at different scales]{Effective field theories at different scales}\label{Sec:SMEFT}

In this section we describe the effective field theory approach to $t\bar t \gamma$ production and \mbox{$b\rightarrow s\gamma$} transitions, for which a set of common dimension-six operators exists. In Sec. \ref{Sec:SMEFT_Top} we give the SMEFT operators considered in our analysis. In Sec. \ref{Sec:WET} we introduce the effective theory for $b\rightarrow s\gamma$ transitions. 

\subsection{Effective Lagrangian for $t\bar{t}\gamma$ production}
\label{Sec:SMEFT_Top}
The effects of heavy BSM particles with mass scale $\Lambda$ can be described at lower energies $E \ll \Lambda$ in a basis of effective operators with mass dimension $d>4$ 
\cite{Weinberg:1978kz,Buchmuller:1985jz}. 
Such higher-dimensional operators are constructed from SM fields and are required to be Lorentz invariant and in accord with SM gauge symmetries.  
The SMEFT Lagrangian $\mathcal{L}_\textmd{SMEFT}$ is an expansion in powers of $\Lambda^{-1}$. 
Higher-dimensional operators $O_i^{(d)}$ of dimension $d$ are added to the SM Lagrangian together with the corresponding Wilson coefficients $C_i^{(d)}$ and a factor $\Lambda^{d-4}$. 
The effective Lagrangian reads
\begin{align}
    \mathcal{L}_\textmd{SMEFT}=\mathcal{L}_\textmd{SM}+\sum_i\frac{C^{(6)}_i}{\Lambda^2}O_i^{(6)}+\mathcal{O}\left(\Lambda^{-4}\right)\,.
    \label{Glg:L_eff}
\end{align}
Operators of dimension $d=5$ and $d=7$ are not considered in this work since they violate lepton and baryon number conservation \cite{Degrande:2012wf,Kobach:2016ami}. 
In the following, we only consider operators with mass dimension $d=6$, which are the leading BSM contributions to LHC physics.\par
A complete basis containing 59 independent operators for one generation (2499 for three generations \cite{Alonso:2013hga}) of fer\-mi\-ons is presented in Ref. \cite{Grzadkowski:2010es} in the \emph{Warsaw basis}, which is used in the following. Fortunately, for any class of observables only a small subset of operators has to be considered. 

\begin{figure*}
  \centering
  \includegraphics[width=0.24\textwidth]{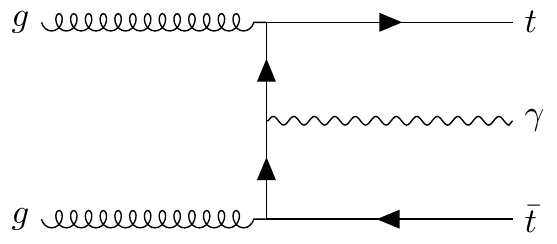}
  \includegraphics[width=0.24\textwidth]{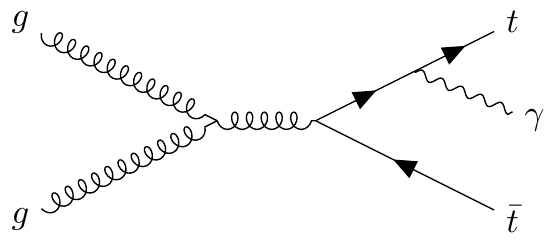}
  \includegraphics[width=0.24\textwidth]{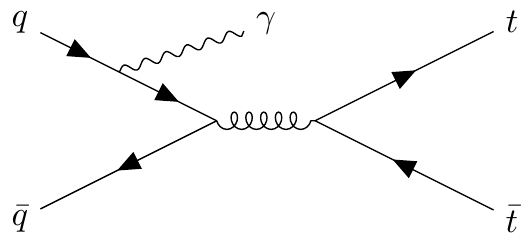}
  \includegraphics[width=0.24\textwidth]{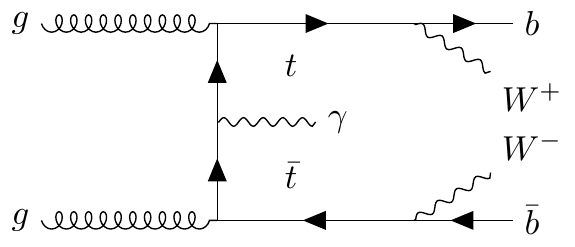}\\
  \vspace{0.3cm}
  \includegraphics[width=0.24\textwidth]{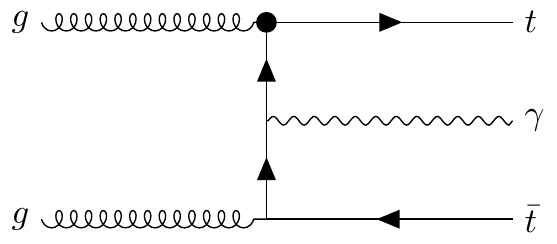}
  \includegraphics[width=0.24\textwidth]{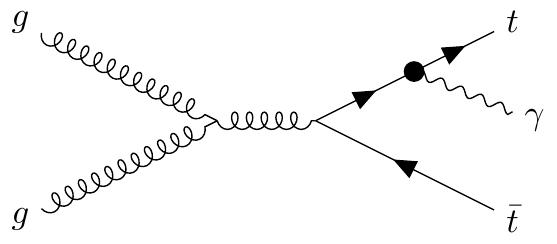}
  \includegraphics[width=0.24\textwidth]{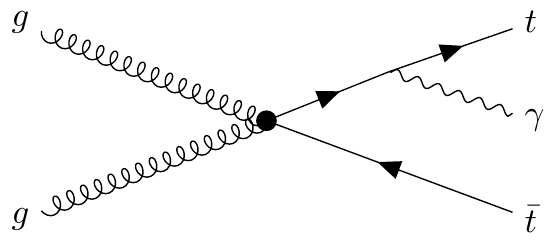}
  \includegraphics[width=0.24\textwidth]{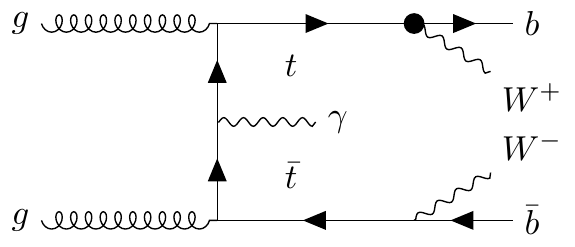}
  \caption{Examples for Feynman diagrams contributing to $t\bar{t}\gamma$ production in $pp$-collisions in the SM (top) and including dimension-six operators (bottom). The black dot denotes the insertion of an effective operator from Eq. (\ref{Eq:TopOperator})}
  \label{Fig:Feynman_ttbary}
\end{figure*}
We study the dimension-six operators affecting $t\bar{t}\gamma$ production at the LHC. Examples for lowest order Feynman diagrams with both gluons and quarks as initial states are shown in Fig. \ref{Fig:Feynman_ttbary}. 
We consider only operators involving third-generation quarks and bosonic fields, including the Higgs field. 
The corresponding operators can be written as
\begin{align}
    \begin{aligned}
      O_{uB}&=\left(\bar{q}_L\sigma^{\mu\nu}u_R\right)\tilde{\varphi}B_{\mu\nu}	\,, \\
      O_{uG}&=\left(\bar{q}_L\sigma^{\mu\nu}T^{A}u_R\right)\tilde{\varphi}G_{\mu\nu}^{A}	\,, \\
      O_{uW}&=\left(\bar{q}_L\sigma^{\mu\nu}\tau^{I}u_R\right)\tilde{\varphi}W_{\mu\nu}^{I}	\,,
      \label{Eq:TopOperator}
      \end{aligned}
\end{align}
with $q_L$ the $SU(2)$ doublet, $u_R$ the up-type $SU(2)$ singlet, the gauge field strength tensors $B_{\mu\nu}$, $W^I_{\mu\nu}$ and $G^A_{\mu\nu}$ of $U(1)_Y$, $SU(2)_L$ and $SU(3)_C$ and the generators $T^A$ and $\tau^I$ of $SU(3)_C$ and $SU(2)_L$, respectively. 
The Higgs-doublet is denoted by $\varphi$ and $\tilde \varphi = i\tau^2\varphi^*$.
Contributions from dipole operators with right-handed $b$ quarks, which contribute to top-quark decay and via one-loop diagrams to $b\rightarrow s$ transitions,
are suppressed by a factor $m_b/m_t$ relative to the ones with right-handed top quarks, and therefore neglected.  
Generally, the effective operators in Eq. (\ref{Eq:TopOperator}) are non-hermitian which leads to complex-valued Wilson coefficients. 
In this analysis, we assume all Wilson coefficients to be real valued. 
Four-quark operators can in principle also affect $t\bar{t}\gamma$ production.
As $t\bar t$ production at the LHC is dominated by the $gg$ channel ($\sim 75\,\%$ and $\sim90\,\%$ at $8\,\si{\tera\electronvolt}$ and $13\,\si{\tera\electronvolt}$, respectively \cite{Buckley:2015lku}), we neglect contributions from four-quark operators. 
We allow for BSM effects in top-quark decay via $O_{uW}$, see Fig.~\ref{Fig:Feynman_ttbary}.
\subsection{Effective Lagrangian for $\bar B \rightarrow X_s \gamma $ decays} 
\label{Sec:WET}
Rare $b\rightarrow s\gamma$ processes can be described by the Weak Effective Field Theory (WET) Lagrangian \cite{Chetyrkin:1996vx}
\begin{equation}
  \mathcal{L}_\text{WET}=\frac{4G_F}{\sqrt{2}}V_{ts}^ *V_{tb}\sum_{i=1}^8\bar C_iQ_i\,,
  \label{Eq:WETLagrangian}
\end{equation}
where $V_{ij}$ are elements of the CKM matrix, $G_F$ is the Fermi coupling constant, $Q_i$ are effective operators and $\bar C_i$ are the corresponding Wilson coefficients including both SM and BSM contributions.
The effective operators relevant for the processes considered here are the four-fermion operators 
\begin{equation}
  \begin{aligned}
      Q_1&=(\bar s_L\gamma_\mu T^ac_L)(\bar c_L\gamma^\mu T^a b_L)\,,\\
      Q_2&=(\bar s_L\gamma_\mu c_L)(\bar c_L\gamma^\mu  b_L)\,,\\
      Q_3&=(\bar s_L\gamma_\mu b_L)\sum_q(\bar q\gamma^\mu q)\,,\\
      Q_4&=(\bar s_L\gamma_\mu T^ab_L)\sum_q(\bar q\gamma^\mu T^a q)\,,\\
      Q_5&=(\bar s_L\gamma_\mu\gamma_\nu\gamma_\sigma b_L)\sum_q(\bar q\gamma^\mu\gamma^\nu\gamma^\sigma q)\,,\\
      Q_6&=(\bar s_L\gamma_\mu\gamma_\nu\gamma_\sigma T^ab_L)\sum_q(\bar q\gamma^\mu\gamma^\nu\gamma^\sigma T^aq)\,,
  \end{aligned}
\end{equation}
as well as the dipole operators
\begin{equation}
\begin{aligned}
Q_7&=\frac{e}{16\pi^2}m_b(\bar s_L\sigma^{\mu\nu}b_R)F_{\mu\nu}\,,\\
Q_8&=\frac{g_s}{16\pi^2}m_b(\bar s_L\sigma^{\mu\nu}T^ab_R)G^a_{\mu\nu}\,,
\end{aligned}
\end{equation}
with chiral left (right) projectors $L$ ($R$) and the field strength tensor of the photon $F_{\mu\nu}$. We neglect contributions proportional to the small CKM matrix element $V_{ub}$ and to the strange-quark mass. 
\section{Matching at one-loop level}
\label{Sec:Match}
\begin{figure*}
    \centering
    \includegraphics[width=0.8\textwidth]{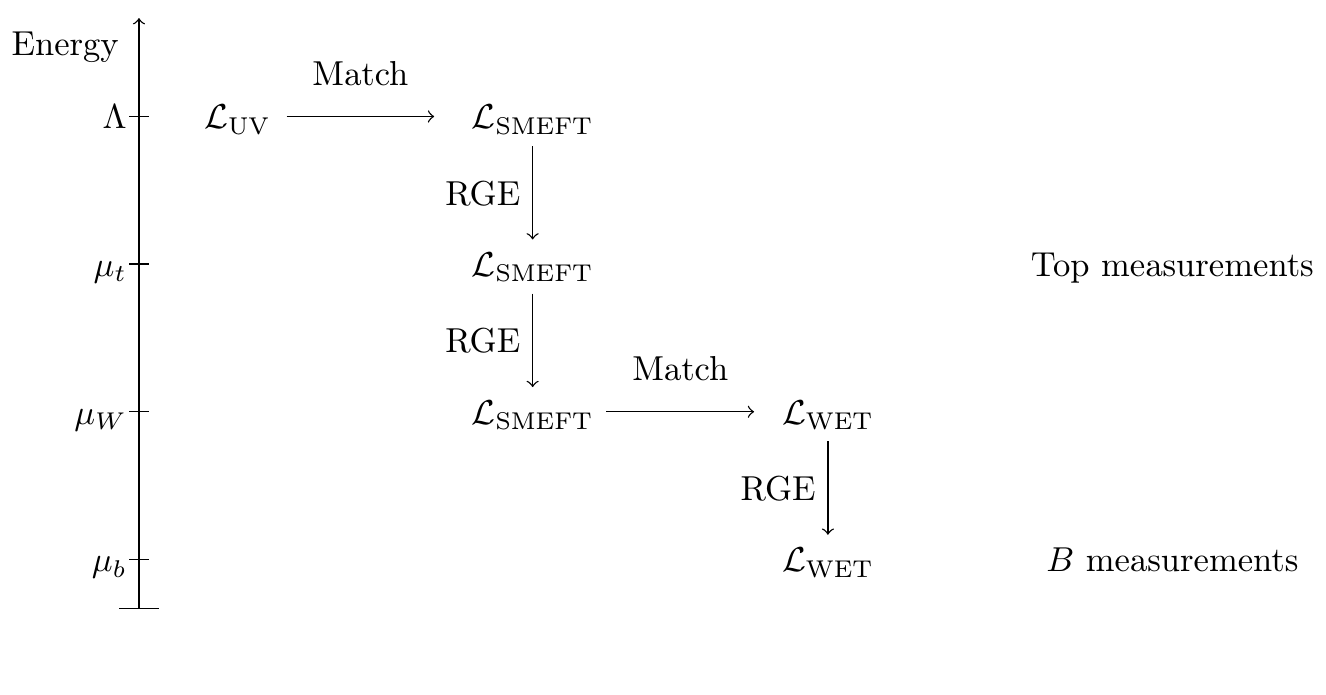}
    \caption{Illustration of the energy scales and effective theories. At the high energy scale $\Lambda$, the UV theory is matched onto SMEFT. For measurements of the top quark, the dimension-six Wilson coefficients in $\mathcal{L}_\textmd{SMEFT}$ are evolved to the scale $\mu_t\sim m_t$ using the SMEFT RGE. For comparison with measurements of $B$ physics, SMEFT is matched at a scale $\mu_W\sim m_W$ onto WET. 
    For measurements at the scale $\mu_b\sim m_b$, the coefficients in $\mathcal{L}_\textmd{WET}$ are evolved using the WET RGE}
    \label{fig:energyscale}
\end{figure*}
To describe BSM physics at energies below the electroweak scale $\mu_W$, the SMEFT Lagrangian in Eq. (\ref{Glg:L_eff}) has to be matched onto the WET Lagrangian as illustrated in Fig. \ref{fig:energyscale}.  Top-quark measurements allow to constrain the values of Wilson coefficients at the scale $\mu_t\sim m_t$.
At the scale $\mu_b\sim m_b$, $B$ measurements can be used to constrain the values of the WET coefficients.
To express $B$ observables in terms of SMEFT Wilson coefficients at the scale $\mu_t$, the following steps 
have to be performed, extending the procedure described in Ref.~\cite{Aebischer:2015fzz}: 
First, RGE evolution of the SMEFT Wilson coefficients from the 
scale $\mu_t$ to $\mu_W$ has to be performed. 
As a next step, $\mathcal{L}_\textmd{SMEFT}$  has to be matched onto $\mathcal{L}_\textmd{WET}$. Finally, the RGE evolution of the WET Wilson coefficients from $\mu_W$ to $\mu_b$ has to be carried out.
These three steps allow the computation of observables, such as BR($\bar B\rightarrow X_s\gamma$), at the scale $\mu_b$ in dependence of the SMEFT Wilson coefficients $C_i(\mu_t)$ at the scale $\mu_t$. In the following, we describe each of the three steps for the $b \rightarrow s \gamma $ process considered in this work. 
\subsection{RGE evolution in SMEFT}
The computation of the RGEs in SMEFT is based on Refs. \cite{Alonso:2013hga,Jenkins:2013zja,Jenkins:2013wua}. To describe the RGE evolution of the operators in Eq. (\ref{Eq:TopOperator}) at $\mathcal{O}(\alpha_s)$, the following SMEFT operators 
have to be included due to mixing:
\begin{equation}
    \begin{aligned}
        O_{u\varphi}&=\left(\varphi^\dagger\varphi\right)\left(\bar q_L u_R \tilde \varphi\right)\,,\\
        O_{\varphi G}&=\left(\varphi^\dagger\varphi\right)G^A_{\mu\nu}G^{A\mu\nu}\,,
        \\
        O_{\varphi \tilde G}&=\left(\varphi^\dagger\varphi\right)\tilde G^A_{\mu\nu}G^{A\mu\nu}\,,\quad
    \end{aligned}
\end{equation}
with $\tilde G^A_{\mu\nu} = \frac{1}{2}\epsilon_{\mu\nu\alpha\beta}G^{A\alpha\beta}$ ($\epsilon_{0123}=+1$). To compute the a\-nom\-alous dimension matrix at $\mathcal{O}(\alpha_s)$, the effective operators have to be rescaled \cite{Jenkins:2013sda}:
\begin{equation}
    \begin{aligned}
        O^\prime_{uB}&= yg^\prime\left(\bar{q}_L\sigma^{\mu\nu}u_R\right)\tilde{\varphi}B_{\mu\nu}	\,, \\
        O^\prime_{u\varphi}&=y\left(\varphi^\dagger\varphi\right)\left(\bar q_L u_R \tilde \varphi\right)\,,\\ 
        O^\prime_{uG}&=yg_s\left(\bar{q}_L\sigma^{\mu\nu}T^{A}u_R\right)\tilde{\varphi}G_{\mu\nu}^{A}	\,, \\
        O^\prime_{\varphi G}&=g_s^2\left(\varphi^\dagger\varphi\right)G^A_{\mu\nu}G^{A\mu\nu}\,,\\
        O^\prime_{uW}&=yg\left(\bar{q}_L\sigma^{\mu\nu}\tau^{I}u_R\right)\tilde{\varphi}W_{\mu\nu}^{I}	\,,\\
        O^{\prime}_{\varphi \tilde G}&=g_s^2\left(\varphi^\dagger\varphi\right)\tilde G^{A}_{\mu\nu}G^{A\mu\nu}\,,
    \end{aligned}
\end{equation}
where $g^\prime$, $g$ and $g_s$ are the coupling constants corresponding to $U(1)_Y$, $SU(2)_L$ and $SU(3)_C$, respectively, and $y$ denotes a Yukawa coupling. The Wilson coefficients change with inverse powers of the couplings. In terms of the rescaled coefficients, the RGEs in SMEFT read 
\begin{equation}
    \frac{d}{d\ln\mu}\begin{pmatrix}
        C_{uG}^\prime\\C_{uW}^\prime\\C_{uB}^\prime\\C^\prime_{u\varphi}\\C^\prime_{\varphi G}\\C^{\prime}_{\varphi \tilde G}
    \end{pmatrix}
    =\frac{\alpha_s}{4\pi} \frac{4}{3}\begin{pmatrix}
        1   &   0   &   0   &   0   &   -3   &   -3i\\
        2   &   2   &   0   &   0   &   0   &   0   \\
        \frac{10}{3}    &   0   &   2   &   0   &   0   &   0\\
        -24   &   0  &    0    &   -6  &  0   &   0\\
        0   &   0   &   0   &   0   &   0   &   0\\
        0   &   0   &   0   &   0   &   0   &   0
    \end{pmatrix}\begin{pmatrix}
        C_{uG}^\prime\\C_{uW}^\prime\\C_{uB}^\prime\\C^\prime_{u\varphi}\\C^\prime_{\varphi G}\\C^{\prime}_{\varphi \tilde G}
    \end{pmatrix}
    \,.
    \label{Eq:SMEFTRGE}
\end{equation}

This matrix is not closed at $\mathcal{O}(\alpha_s)$:
The operators $O^\prime_{\varphi G}$ and 
$O^\prime_{\varphi \tilde G}$ give contributions to the running of
\begin{align}
    O^\prime_{dG}=yg_s\left(\bar{q}_L\sigma^{\mu\nu}T^{A}d_R\right){\varphi}G_{\mu\nu}^{A}
\end{align}
and $O^\prime_{u G}$ contributes to the running of the four-quark operators
\begin{align}
    O^{\prime(1)}_{quqd}&=(q_L^iu_R)\epsilon_{ij}(q_L^jd_R)\,,\\ O^{\prime(8)}_{quqd}&=(q_L^iT^Au_R)\epsilon_{ij}(q_L^jT^Ad_R)\,,
\end{align} 
where $i,j$ are isospin indices and $\epsilon_{12}=+1$. These contributions are suppressed by small down-type Yukawa couplings and neglected in Eq.~(\ref{Eq:SMEFTRGE}). Further more, we see from Eq.~(\ref{Eq:SMEFTRGE}) that $C^\prime_{\varphi G}$ and $C^{\prime}_{\varphi \tilde G}$ do not change their values due to running at $\mathcal{O}(\alpha_s)$. Since $O_{\varphi  G}$ and $O_{\varphi \tilde G}$ have no sizable effect on $t\bar t \gamma$ production \cite{Buckley:2015lku} and $b\rightarrow s\gamma $ transitions, we neglect $O^\prime_{\varphi G}$ and $O^\prime_{\varphi \tilde G}$ under the assumption that only operators including the top quark are generated at the scale $\Lambda$. The operator $ O^\prime_{u\varphi}$ does not directly affect the observables we study but is needed to absorb the UV divergence in the top-quark mass corrections from $O^\prime_{uG}$ in SMEFT NLO computations \cite{Zhang:2014rja}. We compute the BSM contributions at LO QCD and neglect $ O^\prime_{u\varphi}$.
\subsection{Matching SMEFT onto WET}
\begin{figure}[htb]
    \includegraphics[width=0.23\textwidth]{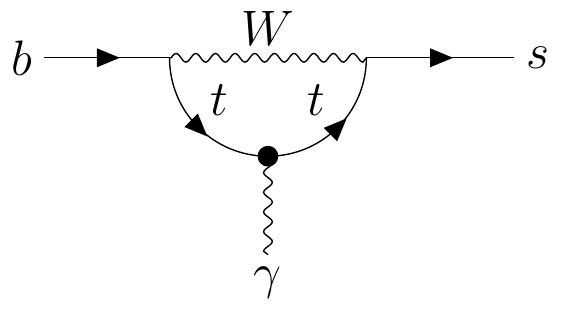}
    \includegraphics[width=0.23\textwidth]{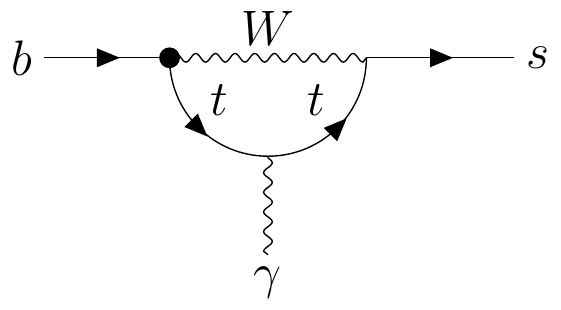}
    \includegraphics[width=0.23\textwidth]{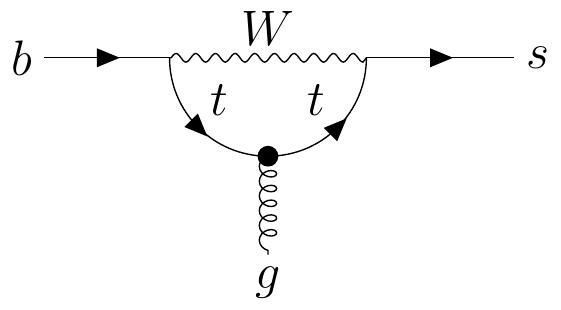}
    \includegraphics[width=0.23\textwidth]{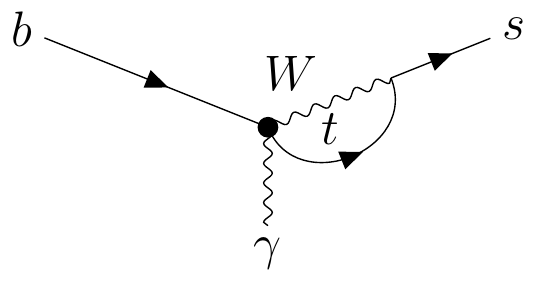}
    \caption{Examples of one-loop diagrams for $b\rightarrow s\gamma$ and $b\rightarrow sg$ transitions. The black dots denote the insertion of a SMEFT operator}
    \label{Fig:MatchingSMEFTWET}
\end{figure}
In Fig. \ref{Fig:MatchingSMEFTWET} we give examples for one-loop diagrams including contributions from  operators in Eq.~(\ref{Eq:TopOperator}) to $\mathcal{L}_\text{WET}$. The matching conditions 
have been calculated in Ref. \cite{Aebischer:2015fzz} and read
\begin{align}
    \begin{aligned}
    \Delta \bar C_{7}^{(0)}&=\frac{\sqrt{2}m_t}{m_W}\bigg[\tilde C_{uW}E_{7}^{uW}(x_t)+\tilde C_{uW}^*F_{7}^{uW}(x_t)
    \\&+\frac{\cos\theta_w}{\sin\theta_w}\left(\tilde C_{uB}E_{7}^{uB}(x_t)+\tilde C_{uB}^*F_{7}^{uB}(x_t)\right)\bigg]\,,    \label{Eq:MatchC7}
    \end{aligned}
\end{align}
\begin{align}
    \begin{aligned}
    \Delta \bar C_{8}^{(0)}&=\frac{\sqrt{2}m_t}{m_W}\bigg[\tilde C_{uW}E_{8}^{uW}(x_t)+\tilde C_{uW}^*F_{8}^{uW}(x_t)\\
    &-\frac{g}{g_s}\left(\tilde C_{uG}E_8^{uG}(x_t)+\tilde C_{uG}^*F_8^{uG}(x_t)\right)\bigg]\,,
    \label{Eq:MatchC8}
    \end{aligned}
\end{align}
where $x_t=m_t^2/m_W^2$ and $\Delta \bar C_i^{(0)}$ denotes BSM contributions at order $\alpha_s^0$ to the coefficients in $\mathcal{L}_{\textmd{WET}}$. The $\tilde C_i$ denote rescaled Wilson coefficients 
\begin{equation}
    \tilde C_i =C_i \frac{v^2}{\Lambda^2}\,,
\end{equation}
where $v=\SI{246}{\GeV}$ is the Higgs vacuum expectation value.
Explicit expressions for the $x_t$-dependent functions $E_{7}^{uW}$, $F_{7}^{uW}$, $E_{8}^{uW}$ and $F_{8}^{uW}$ can be found in Ref. \cite{Aebischer:2015fzz} and are given in \ref{App:Match}.\par
\subsection{RGE evolution in WET}
At the scale $\mu_W$, both the SM and BSM contributions are matched onto $\mathcal{L}_\textmd{WET}$. The RGEs are then used to evolve the coefficients $\bar C_i$ from $\mu_W$ to $\mu_b$. By doing so, large logarithms are resummed to all orders in perturbation theory. Instead of the original coefficients $\bar C_i$
it is convenient to use the effective coefficients \cite{Buras:1993xp,Greub:1996jd}
\begin{equation}
    C_i^\textmd{eff}=\begin{cases}
        \bar{C}_i & \textmd{for }i=1,...,6\\
        \bar C_7+\sum_{j=1}^6y_j\bar C_j & \textmd{for }i=7\\
        \bar C_8+\sum_{j=1}^6z_j\bar C_j & \textmd{for }i=8\\
    \end{cases}\,.
\end{equation}
One finds $y=(0,0,-1/3,-4/9,-20/3,-80/9)$ and $z=(0,0,1,-1/6,20,-10/3)$ \cite{Chetyrkin:1996vx} in the $\overline{MS}$ scheme with fully anticommuting $\gamma_5$. 
The RGEs for the effective coefficients read
\begin{equation}
    \frac{d}{d\ln\mu}C_i^\textmd{eff}(\mu)= \gamma^\textmd{eff}_{ji}(\mu)C_j^\textmd{eff}(\mu)\,,
    \label{Eq:WETRGE}
\end{equation}
with the anomalous dimension matrix $\gamma^\textmd{eff}$. The perturbative expansion of this matrix is given as
\begin{equation}
    \gamma^\textmd{eff}(\mu)=\frac{\alpha_s(\mu)}{4\pi}\gamma^{(0)\textmd{eff}}+\frac{\alpha_s^2(\mu)}{(4\pi)^2}\gamma^{(1)\textmd{eff}}+\frac{\alpha_s^3(\mu)}{(4\pi)^3}\gamma^{(2)\textmd{eff}}+...\ .
    \label{Eq:ADMWET}
\end{equation}
The matrices $\gamma^{(0)\textmd{eff}}$ and $\gamma^{(1)\textmd{eff}}$ are given in Ref. \cite{Chetyrkin:1996vx}. The matrix $\gamma^{(2)\textmd{eff}}$ is specified in Ref. \cite{Czakon:2006ss}. 
Analogously, the coefficients expanded in powers of $\alpha_s$ read 
\begin{align}
    \begin{aligned}
    C_i^\textmd{eff}(\mu)&=C_i^{(0)\textmd{eff}}(\mu)+\frac{\alpha_s(\mu)}{4\pi}C_i^{(1)\textmd{eff}}(\mu)\\&+\frac{\alpha_s^2(\mu)}{(4\pi)^2}C_i^{(2)\textmd{eff}}(\mu)+...\ .
    \end{aligned}
\end{align}
The SM values of the effective coefficients at the scale $\mu_W$ are known at NNLO QCD \cite{Czakon:2015exa,Bobeth:1999mk,Misiak:2004ew}. \par
Obviously, performing the matching of $\tilde C_i$ to $\Delta \bar C^{(0)}_i$ without running in SMEFT and WET only by setting $\mu_W=\mu_b$ in Eq.~(\ref{Eq:MatchC7}) and Eq.~(\ref{Eq:MatchC8}) leads to a completely different dependence of the SMEFT coefficients. The impact of the $\tilde C_i$ on $\Delta \bar C^{(0)}_i$ can become larger by factors up to $\approx 40$ and contributions due to mixing are not included. 
\section{Measurements}\label{Sec:Measurements}
In this section, the measurements of the $t\bar t \gamma$ production cross section and of the $\bar B \rightarrow X_s \gamma $ branching fraction that we use for constraining the Wilson coefficients are described.

\subsection{Measurements of the $t\bar{t}\gamma$ cross section}\label{tta_measurements}
Cross sections of $t\bar{t}\gamma$ production have been measured at different center-of-mass energies by the ATLAS \cite{Aad:2015uwa, Aaboud:2017era, ATLAS_13} and CMS \cite{Sirunyan:2017iyh} experiments. For our fits, we consider the cross sections determined in the \SI{13}{\TeV} analysis performed by the ATLAS collaboration using 2015 and 2016 LHC data corresponding to an integrated luminosity of \SI{36.1}{\per\femto\barn} \cite{ATLAS_13}. In this \mbox{analysis}, the $t\bar{t}\gamma$ production cross section is reported as a fiducial cross section for final states containing one or two leptons (in the following referred to as single-lepton or dilepton channel, respectively), where the leptons can be either electrons or muons (or their corresponding antiparticles). The fiducial regions for both channels are defined in Sec. 7.1 of Ref. \cite{ATLAS_13}.
The measured values of the single-lepton and dilepton fiducial cross sections are reported as
\begin{align*}
    \sigma^\mathrm{fid}_\mathrm{ATLAS}(t\bar{t}\gamma, 1\ell) &= 521 \pm 9 \, \text{(stat.)} \pm 41 \, \text{(syst.)}\, \si{\femto\barn}\,,\nonumber\\
    \sigma^\mathrm{fid}_\mathrm{ATLAS}(t\bar{t}\gamma, 2\ell) &= 69 \pm 3 \, \text{(stat.)} \pm 4 \, \text{(syst.)}\, \si{\femto\barn}\, .\nonumber
\end{align*}
Within uncertainties, the measurements agree well with the SM predictions at NLO QCD \cite{ATLAS_13,Melnikov:2011ta}: 
\begin{align*}
    \sigma^\mathrm{fid}_\mathrm{SM, NLO}(t\bar{t}\gamma, 1\ell) &= 495 \pm \SI{99}{\femto\barn}\,,\nonumber\\
    \sigma^\mathrm{fid}_\mathrm{SM, NLO}(t\bar{t}\gamma, 2\ell) &= 63 \pm \SI{9}{\femto\barn}\, .\nonumber
\end{align*}
\subsection{Measurements of BR($ \bar B \rightarrow X_s \gamma $)} \label{Sec:Meas_B}
For the branching fraction of $\bar B \rightarrow X_s \gamma $ multiple measurements, performed by the BaBar \cite{Aubert:2007my, Lees:2012ym, Lees:2012wg}, Belle \cite{Limosani:2009qg, Saito:2014das, Belle:2016ufb} and CLEO \cite{Chen:2001fja} experiments, are available. 
A combination of these measurements has been performed by the Heavy Flavor Averaging Group (HFLAV) \cite{HFLAV16}, taking into account the different minimum photon energy requirements applied in the respective analyses. 
The differences are corrected for by performing an extrapolation according to the method described in Ref. \cite{Buchmuller:2005zv}.
For our fits we use the most recent result of the combination of $\text{BR}(\bar B \rightarrow X_s \gamma)$ measurements \cite{HFLAV_19}, 
\begin{align*}
    \text{BR}(\bar B \rightarrow X_s \gamma) = (332 \pm 15)\times 10^{-6}\,,
\end{align*}
with a minimum photon energy requirement of $E_\gamma >\SI{1.6}{\GeV}$. This value agrees well with the NNLO SM prediction \cite{Misiak:2015xwa}
\begin{align*}
    \text{BR}_\text{SM}(\bar B \rightarrow X_s \gamma) = (336 \pm 23)\times 10^{-6}\,.
\end{align*}

\section{Modeling observables}
\label{Sec:Modelling}
In the following we describe the computation of the SM and BSM contributions to the observables. In Sec. \ref{Sec:CStty} we discuss how to model the fiducial $t\bar{t}\gamma$ cross section and in Sec. \ref{Sec:BRBXsy} we describe the computation of $\text{BR} (\bar B \rightarrow X_s \gamma )$.

\subsection{Computation of $\sigma(t\bar t\gamma)$}
\label{Sec:CStty}
The $t\bar{t}\gamma$ production cross section can be computed at LO QCD for any given configuration of Wilson coefficients using Monte Carlo (MC) simulations.
Since the MC simulations take too long to be directly interfaced to the fit of Wilson coefficients, we determine a parametrization of $\sigma(t\bar t\gamma)$ in terms of the Wilson coefficients.
By squaring the matrix element of processes including dimension-six operators, the cross section in the presence of Wilson coefficients $\tilde C_i$ can be expressed as \begin{equation}
    \sigma = \sigma^\mathrm{SM} + \sum_i \tilde C_i\sigma_i^\text{interf.} + \sum_{i \leq j} \tilde C_i \tilde C_j \sigma_{ij}^\text{BSM}\,, 
    \label{Eq:interpol}
\end{equation}
where $\sigma_i^\text{interf.}$ are terms coming from the interference between SM and EFT diagrams and $\sigma_{ij}^\text{BSM}$ are purely BSM contributions.
Using cross sections computed with MC simulations for different configurations of Wilson coefficients as sampling points, an interpolation to Eq. (\ref{Eq:interpol}) can be performed, yielding numerical values for the $\sigma_i$ terms and thus a parametrization of the cross section as a function of the Wilson coefficients that can be used in the fit.
\par
To parametrize the impact of the dimension-six operators $O_{uB}$, $O_{uG}$ and $O_{uW}$ on the $t\bar t\gamma$ production cross section, we perform simulations using \MG{}\cite{MG5} with the \texttt{dim6top\_LO} UFO model \cite{AguilarSaavedra:2018nen}. 
We generate MC samples similar to the signal sample described in Ref. \cite{ATLAS_13} to make sure that the simulations are suitable for a fit to the fiducial measurements. 
The samples are generated using $2 \rightarrow 7$ processes for both, the single-lepton and the dilepton channel, allowing for BSM contributions from $O_{uW}$ in top-quark decay. 
For the BSM contributions only one insertion of a dimension-six operator is allowed  at a time and the BSM energy scale is set to $\Lambda =\SI{1}{\TeV}$.
The dimension-six operators we consider in this paper are $O_{uB}$, $O_{uG}$ and $O_{uW}$, as given in Eq. (\ref{Eq:TopOperator}). 
In the \texttt{dim6top\_LO} UFO model different degrees of freedom are chosen than in this analysis, so that it is not possible to directly specify the value of the coefficient $\tilde C_{uB}$ but only the value of the linear combination
\begin{equation}
    \tilde C_{uZ} = \cos\theta_W \tilde C_{uW} - \sin\theta_W \tilde C_{uB}\,, \label{Eq:ctz}
\end{equation}
where $\theta_W$ is the Weinberg angle (in the notation of Ref. \cite{AguilarSaavedra:2018nen} $C_{tZ}$ is used instead of $C_{uZ}$).
Thus, we generate sampling points in the space of the Wilson coefficients $\tilde C_{uG}$, $\tilde C_{uW}$ and $\tilde C_{uZ}$ and use the equivalent representation in terms of $\tilde C_{uB}$, $\tilde C_{uG}$ and $\tilde C_{uW}$ for determining constraints on the coefficients hereinafter. We choose 201 different sampling points, where up to two Wilson coefficients at a time can take non-zero values. 
For each of the sampling points, \SI{50000}{} events are generated.
Comparing the SM value obtained with the cross section of the LO signal sample described in Ref. \cite{ATLAS_13}, we find good agreement with a relative deviation of less than \SI{4}{\percent}.\par
We determine the parametrization of the $t \bar t \gamma$ cross sections as a function of the Wilson coefficients $\tilde{C}_{uG}\, ,\ \tilde{C}_{uW}$ and $\tilde{C}_{uZ}$ by performing an interpolation according to Eq. (\ref{Eq:interpol}). For the interpolation we apply a least squares fit with the Leven\-berg--Marquardt algorithm provided by the \texttt{LsqFit.jl} package \cite{LsqFit}.\par
\begin{figure*}
    \centering
    \includegraphics[width=0.325\textwidth]{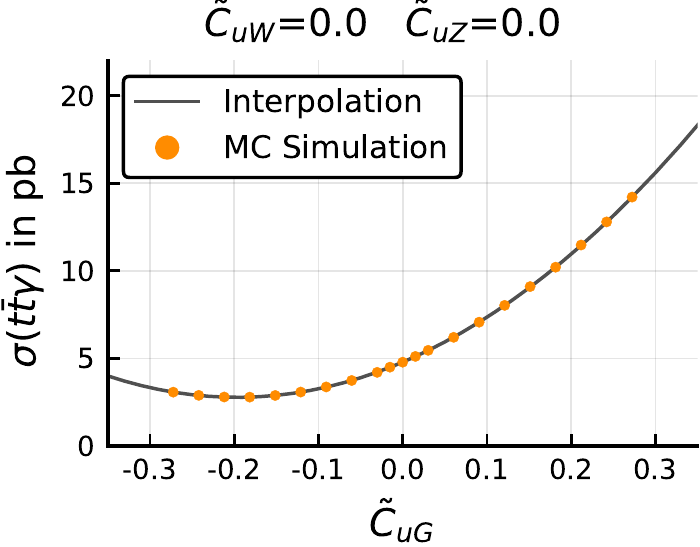}
    \includegraphics[width=0.325\textwidth]{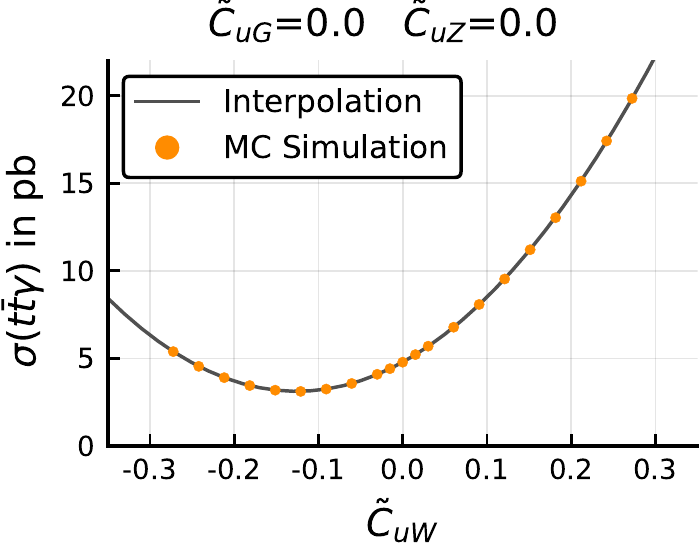}
    \includegraphics[width=0.325\textwidth]{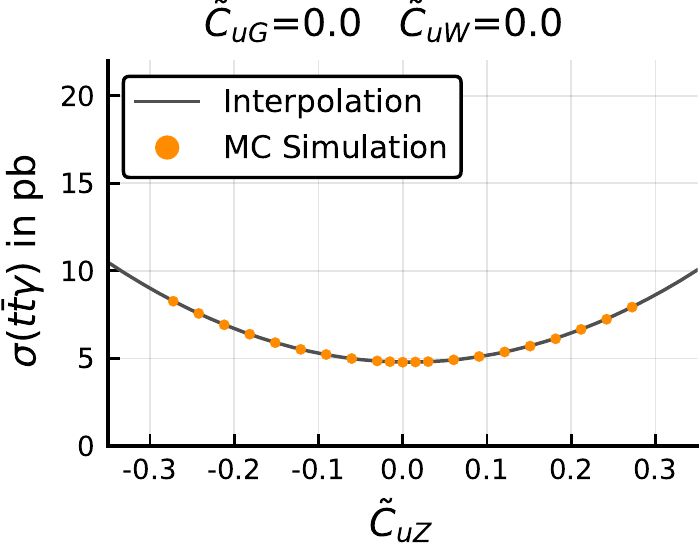}
    \caption{Sampling points and interpolation result for the $t \bar t \gamma$ cross section, represented as slices of the phase space where only one of the Wilson coefficient is varied at a time, while the others are set to zero}
    \label{Fig:Interpolation_total}
\end{figure*}

The sampling points and the result of the interpolation are shown in Fig. \ref{Fig:Interpolation_total} as slices of the phase space where only one Wilson coefficient is varied at a time, while the others are set to zero. 
We find that the simulated cross sections are well described by the interpolation, as the relative differences between the simulated values and the interpolation, calculated at all sampling points, have a standard deviation of only \SI{0.2}{\percent}.
\begin{figure*}
    \centering
    \includegraphics[width=0.325\textwidth]{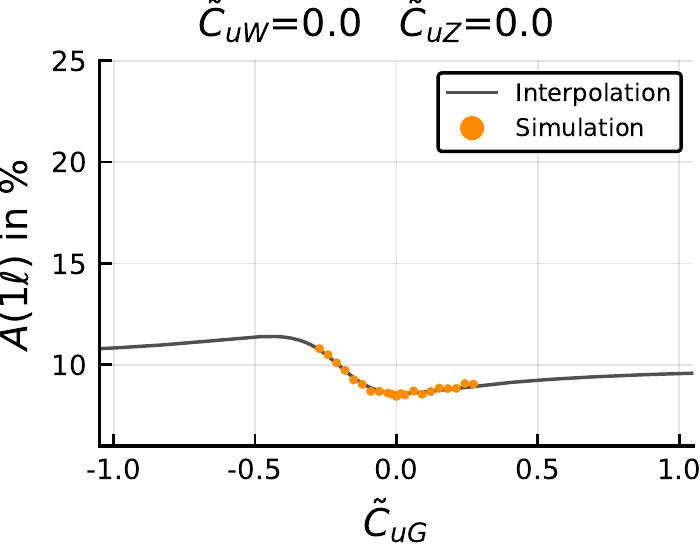}
    \includegraphics[width=0.325\textwidth]{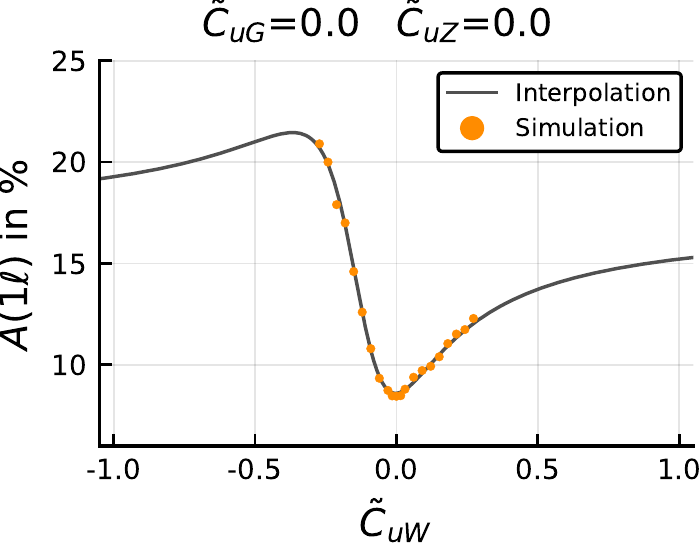}
    \includegraphics[width=0.325\textwidth]{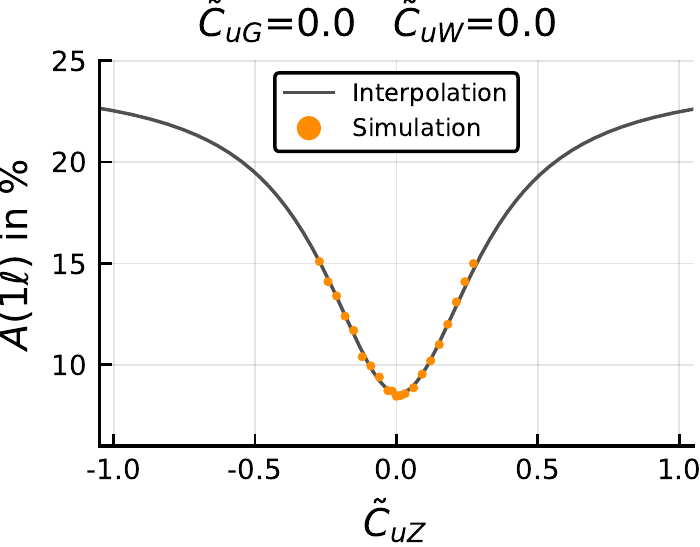}
    \caption{Sampling points and interpolation result for the fiducial acceptance of the single-lepton channel $A(1\ell)$, represented as slices of the phase space where only one of the Wilson coefficient is varied at a time, while the others are set to zero}
    \label{Fig:acceptance}
\end{figure*}
To obtain fiducial acceptances, we apply parton showering to the events using \texttt{PYTHIA8} \cite{PYTHIA8} and perform a particle-level event selection with \texttt{MadAnalysis} \cite{Conte:2012fm, Conte:2014zja, Dumont:2014tja}. For the clustering of particle jets, the anti-$k_t$ algorithm \cite{Cacciari:2008gp} with a radius parameter $R=0.4$ is applied using \texttt{FastJet} \cite{Cacciari:2011ma}.
At each sampling point we determine the fiducial acceptances for the single-lepton and dilepton channels using an event selection that is similar to the definition of the fiducial regions described in Ref. \cite{ATLAS_13}. 
Comparisons of the fiducial acceptances for the SM sampling point with the values given in Ref. \cite{ATLAS_13} show that we obtain the same fiducial acceptance for the dilepton channel and only a small deviation of \SI{3}{\percent} for the single-lepton channel.
\par
It should be noted that performing a parton-level simulation and applying the fiducial cuts at this level, which might be considered as a first approximation, is not sufficient as the resulting LO fiducial cross sections deviate from the LO SM predictions in Ref. \cite{ATLAS_13} by about \SI{50}{\percent} for the single-lepton and \SI{25}{\percent} for the dilepton channel.\par 
The dependence of the fiducial acceptance $A$ on the Wilson coefficients $\tilde C_i$ can be parametrized as
\begin{equation}
    A = \frac{A^\mathrm{SM}\sigma^\mathrm{SM} + \sum_i \tilde{C}_i A_i^\mathrm{interf.} \sigma_i^\mathrm{interf.}  + \sum_{i \leq j} \tilde C_i \tilde C_j A_{ij}^\text{BSM} \sigma_{ij}^\text{BSM}}{\sigma^\mathrm{SM} + \sum_i \tilde C_i\sigma_i^\text{interf.} + \sum_{i \leq j} \tilde C_i \tilde C_j \sigma_{ij}^\text{BSM}}\,,
    \label{Eq:acceptance}
\end{equation} 
where the denominator is the  parametrization of the cross section $\sigma$ as given in Eq. (\ref{Eq:interpol}).
The acceptances $A_i$  account for changes in kinematics due to BSM contributions.
With the parameters $\sigma_i$ already determined in the previous interpolation of the cross section, we perform a least squares fit of the fiducial acceptances to Eq. (\ref{Eq:acceptance}) in each channel using the acceptances from the event selection as sampling points.
The result of the interpolation and the sampling points for the fiducial acceptance of the single-lepton channel are shown in Fig. \ref{Fig:acceptance}.
It is observable that the Wilson coefficients $\tilde{C}_{uW}$ and $\tilde{C}_{uZ}$ have a stronger impact on the acceptance than $\tilde{C}_{uG}$. Compared to the SM value, the former coefficients can both change the acceptance by up to a factor of 2.5, while the latter changes it only by up to a factor of 1.3. 
For the fiducial acceptances of the dilepton channel a comparable behavior can be observed. The corresponding plots are shown in \ref{Sec:app_acc}.
In both channels, fluctuations in the simulated acceptances are present. The standard deviation of the relative difference between simulation and interpolation is \SI{1.3}{\percent} in the single-lepton channel and  \SI{3.9}{\percent} in the dilepton channel, indicating that both interpolations are sufficient.
\par
We obtain the dependence of the fiducial cross sections on the Wilson coefficients by multiplying the interpolation of the total cross section with the interpolations of the fiducial acceptances.
As our simulations are performed at LO QCD and NLO calculations of the SM fiducial cross sections are available, we apply a SM $k$-factor by setting the SM contributions to the according values of the NLO predictions presented in Sec. \ref{tta_measurements}.
\par
\begin{figure*}
\includegraphics[width=0.49\textwidth]{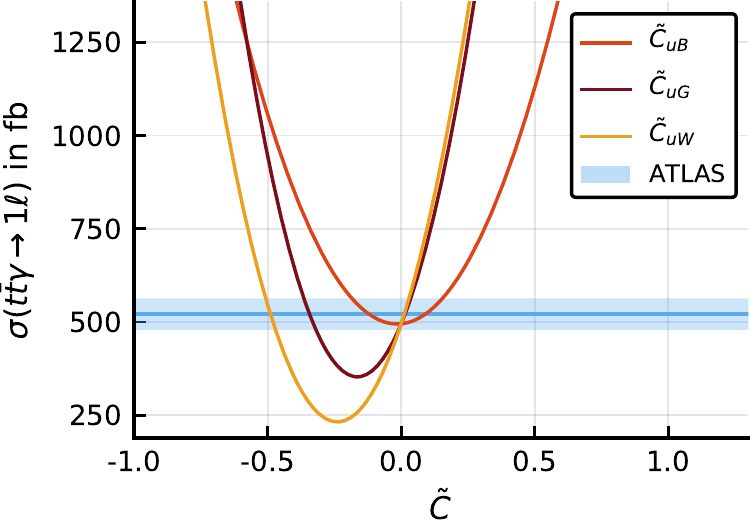}
\hfill
\includegraphics[width=0.49\textwidth]{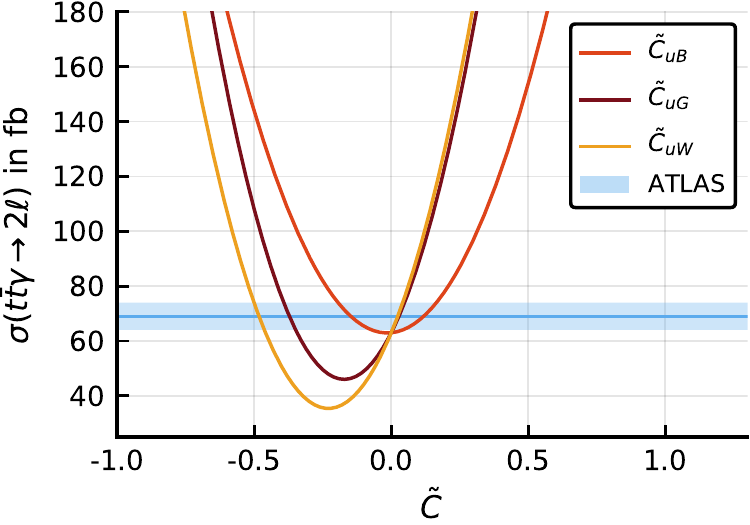}
\caption{Parametrizations of the fiducial $t \bar t \gamma$ cross sections for (left) the single-lepton channel and (right) the dilepton channel, represented as slices of the phase space where only one of the Wilson coefficient is varied at a time, while the others are set to zero. Also indicated are the corresponding ATLAS measurements}
\label{Fig:tta}
\end{figure*}
In Fig. \ref{Fig:tta} the resulting parametrizations of the fiducial $t \bar t \gamma$ cross sections as functions of the Wilson coefficients $\tilde{C}_{uB}\, ,\ \tilde{C}_{uG}$ and $\tilde{C}_{uZ}$ are shown for the single-lepton and dilepton channels. The dependence on $\tilde{C}_{uB}$ is determined using Eq. (\ref{Eq:ctz}). Shown are slices of the phase space where only one Wilson coefficient is varied at a time, while the others are set to zero. 
In both channels, we observe a comparable behavior of the fiducial cross sections and similar sensitivities to the Wilson coefficients. 
\subsection{Computation of BR($\bar B \rightarrow X_s \gamma $)} \label{Sec:BRBXsy}
The most recent estimate of the $\bar B \rightarrow X_s \gamma $ branching fraction at NNLO QCD has been presented in Ref. \cite{Misiak:2015xwa}, following the algorithm described in Ref. \cite{Czakon:2015exa}. We adapt this procedure in our computation of BR($\bar B \rightarrow X_s \gamma $) and extend it to LO BSM contributions.
Applying the notation of Ref. \cite{Misiak:2006ab}, the branching fraction can be expressed as
\begin{align}
\begin{aligned}
    \textmd{BR}(\bar B \rightarrow X_s \gamma)=&\textmd{BR}(\bar B \rightarrow X_c e\bar\nu)_\textmd{exp} \\ 
    &\times \left|\frac{V_{ts}^*V_{tb}}{V_{cb}}\right|^2\frac{6\alpha_{e}}{\pi C}(P(E_0)+N(E_0))\,,
    \end{aligned}
\end{align}
where $\alpha_{e}$ is the fine structure constant, $E_0=1.6\,\si{\giga\electronvolt}$ is the photon energy cut and $P(E_0)$ and $N(E_0)$ denote perturbative and non-perturbative corrections, respectively. The factor $C$ is given as 
\begin{equation}
    C=\left|\frac{V_{ub}}{V_{cb}}\right|^2\frac{\Gamma(\bar B \rightarrow X_c e\bar\nu)}{\Gamma(\bar B \rightarrow X_u e\bar\nu)}\,,
\end{equation}
with an experimental value $C_\textmd{exp}=0.568\pm0.007\pm0.01$ \cite{Alberti:2014yda}. The quantity $P(E_0)$ is given as
\begin{equation}
    P(E_0)=\sum_{i,j=1}^8C^\textmd{eff}_i(\mu_b)C^\textmd{eff}_j(\mu_b)K_{ij}(E_0,\mu_b)\,,
    \label{Eq:PE0}
\end{equation}
where the matrix $K(E_0,\mu_b)$ expanded in $\alpha_s$ reads:
\begin{align}
\begin{aligned}
    K_{ij}(E_0,\mu_b)=&\delta_{i7}\delta_{j7}+\frac{\alpha_s(\mu_b)}{4\pi}K^{(1)}_{ij}\\
    &+\frac{\alpha_s^2(\mu_b)}{(4\pi)^2}K^{(2)}_{ij}+\mathcal{O}(\alpha_s^3(\mu_b))\,.
    \label{Eq:Ki_expand}
\end{aligned}
\end{align}
The coefficients $K^{(1)}_{ij}$ can be derived from the NLO results given in Ref. \cite{Buras:2002tp}. 
For the computation of $P(E_0)$ at approximate NNLO we include the effects of charm and bottom masses in $K^{(2)}_{77}$ \cite{Asatrian:2006rq}, $K^{(2)}_{78}$ \cite{Ewerth:2008nv} and $K^{(2)}_{1(2)7}$ \cite{Boughezal:2007ny} as well as the complete computation of $K^{(2)}_{78}$ \cite{Asatrian:2010rq} and the NNLO computation of $K^{(2)}_{1(2)7}$ \cite{Czakon:2015exa}.
Contributions of three-body and four-body final states to $K^{(2)}_{88}$ \cite{Ferroglia:2010xe,Misiak:2010tk} and $K^{(2)}_{1(2)8}$\cite{Misiak:2010tk} are included in the Brodsky--Lepage--Mackenzie (BLM) approximation \cite{Brodsky:1982gc}. 
For the computation of non-perturbative corrections we include results from \cite{Benzke:2010js,Ewerth:2009yr,Alberti:2013kxa}. The scales are chosen to be $\mu_W=m_W$ and $\mu_b=\SI{2}{\giga\electronvolt}$. 
For the SM central value we find BR$_\textmd{SM}(\bar B\rightarrow X_s\gamma)=336\times 10^{-6}$, matching the results in Ref. \cite{Misiak:2015xwa}.\par
\begin{figure}[htb]
\centering
\includegraphics[width=0.49\textwidth]{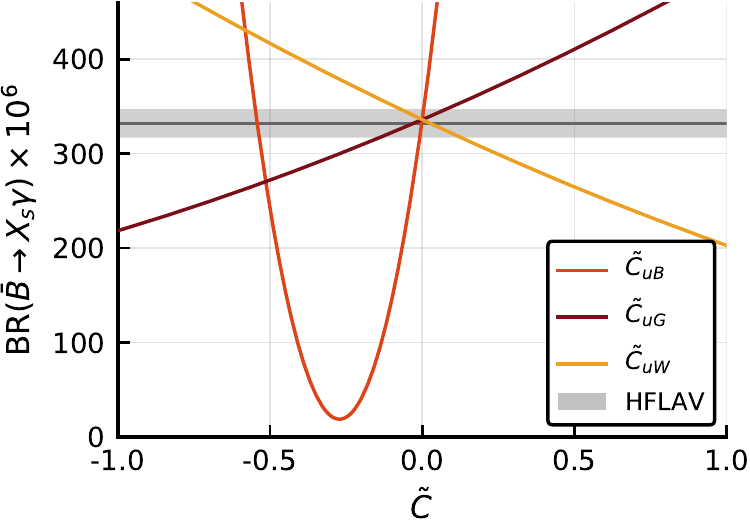}
\caption{Dependence of BR($\bar B \rightarrow X_s \gamma $) on the SMEFT coefficients \mbox{$\tilde C_i(\mu=m_t)$}. Only one coefficient is varied at a time while the other two are set to zero. The grey band denotes the experimental average}
\label{Fig:BRSmeftCoeff}
\end{figure}

 
In Fig. \ref{Fig:BRSmeftCoeff} we give the dependence of BR($\bar B \rightarrow X_s \gamma $) on the SMEFT coefficients at the scale $\mu=m_t$.  
Only one coefficient is varied while the other two are set to zero. 
We also indicate the averaged measurements described in Sec. \ref{Sec:Meas_B}. The branching fraction BR($\bar B \rightarrow X_s \gamma $) shows the strongest dependence on $\tilde C_{uB}$, 
whereas the dependence on $\tilde C_{uG}$ and $\tilde C_{uW}$ is weaker.
Numerically, Eq.~\eqref{Eq:MatchC7} reads for real-valued Wilson coefficients
$\Delta \bar C_{7}^{(0)}(\mu_W)=0.093 \tilde C_{uW}(\mu_W) - 2.354 \tilde C_{uB}(\mu_W)$ and $\tilde C_{uG}$ is of higher order in $\alpha_s$.
As a cross check for our computation, we apply \texttt{flavio} \cite{Straub:2018kue} together with \texttt{wilson} \cite{Aebischer:2018bkb} and Eq. (\ref{Eq:MatchC7}) and Eq. (\ref{Eq:MatchC8}) to compute the branching fraction. 
Since \texttt{wilson} provides only tree-level matching between SMEFT and WET, the matching conditions in Eq. (\ref{Eq:MatchC7}) and Eq. (\ref{Eq:MatchC8}) are not included. 
We therefore apply \texttt{wilson} only for the RGE evolution in WET. 
For the SM prediction we find good agreement with the result obtained using \texttt{flavio}, BR$_\text{flavio}(\bar B \rightarrow X_s \gamma)= (326\pm23)\times 10^{-6}$. The deviation of the central value is only \SI{2}{\percent} and thus smaller than the theory uncertainties.
For the dependence on the Wilson coefficients we find very similar behavior and obtain only deviations smaller than the theory uncertainties in the range $-1\leq \tilde C_i \leq 1$.

\section{Constraining Wilson coefficients}\label{Sec:Fit}
With the parametrizations of the $t\bar t \gamma$ cross sections and of the \mbox{$\bar B \rightarrow X_s\gamma$} branching fraction determined in Sec. \ref{Sec:Modelling}, we perform fits to the measurements described in \mbox{Sec. \ref{Sec:Measurements}} to constrain the Wilson coefficients $\tilde{C}_{uB}$, $\tilde{C}_{uG}$ and $\tilde{C}_{uW}$.
We use a new implementation of the EFT\emph{fitter} tool \cite{Castro:2016jjv} based on the \emph{Bayesian Analysis Toolkit - BAT.jl} \cite{BAT, BAT.jl}. This allows to perform fits of Wilson coefficients in a Bayesian reasoning, yielding (marginalized) posterior probability distributions of the parameters.\par
We include both the experimental uncertainties and the SM theory uncertainties given in Sec. \ref{Sec:Measurements} in the fit.
Focusing on the combination of observables from different energy scales, we make the simplifying assumption that the uncertainties of the measurements included are gaussian distributed \cite{Castro:2016jjv} and uncorrelated. 
This assumption seems reasonable for the correlations between top-quark and $B$ physics measurements and also for the correlation between the statistical uncertainties of the two channels contributing to $\sigma(t\bar t \gamma)$. 
The systematic and theoretical uncertainties of both channels can in principle be correlated in a non-negligible manner. As no information about the correlations is available, we investigate their impact afterwards by performing several fits varying the corresponding correlation coefficients.
\par
To illustrate the benefit of combining observables from top-quark and $B$ physics, we first constrain the Wilson coefficients using only one set of measurements at a time (Secs.~\ref{sec:b}, \ref{sec:t}) before performing the combined fit (Sec.~\ref{sec:tplusb}).
\subsection{$B$ physics only \label{sec:b}}
Considering only BR$(\bar B \rightarrow X_s \gamma)$, we perform a fit to the HFLAV average described in Sec. \ref{Sec:Meas_B} using the description of the branching fraction given in Sec. \ref{Sec:BRBXsy}. 
Treating $\tilde{C}_{uB}$, $\tilde{C}_{uG}$ and $\tilde{C}_{uW}$ as free parameters of the fit and providing no prior knowledge about their distributions, we assign uniform prior probability distributions in the range of [-1, 1] to them.
Larger values of the rescaled Wilson coefficients $\tilde C$ would not be reasonable and would lead to a breakdown of the EFT expansion.\par
When performing the fit, we observe that only $\tilde{C}_{uB}$ can be constrained using this setup. No constraints on the other two coefficients can be obtained, as the resulting marginalized posterior probabilities of $\tilde{C}_{uG}$ and $\tilde{C}_{uW}$ are uniformly distributed. 
As can be seen from Fig. \ref{Fig:BRSmeftCoeff}, $\tilde{C}_{uB}$ is the Wilson coefficient with the largest impact on the $\bar B \rightarrow X_s \gamma $ branching fraction, thus receiving stronger constraints than $\tilde{C}_{uG}$ and $\tilde{C}_{uW}$ in a fit with three free parameters and a single observable.
\par
\begin{figure}
    \centering
    \includegraphics[width=0.49\textwidth]{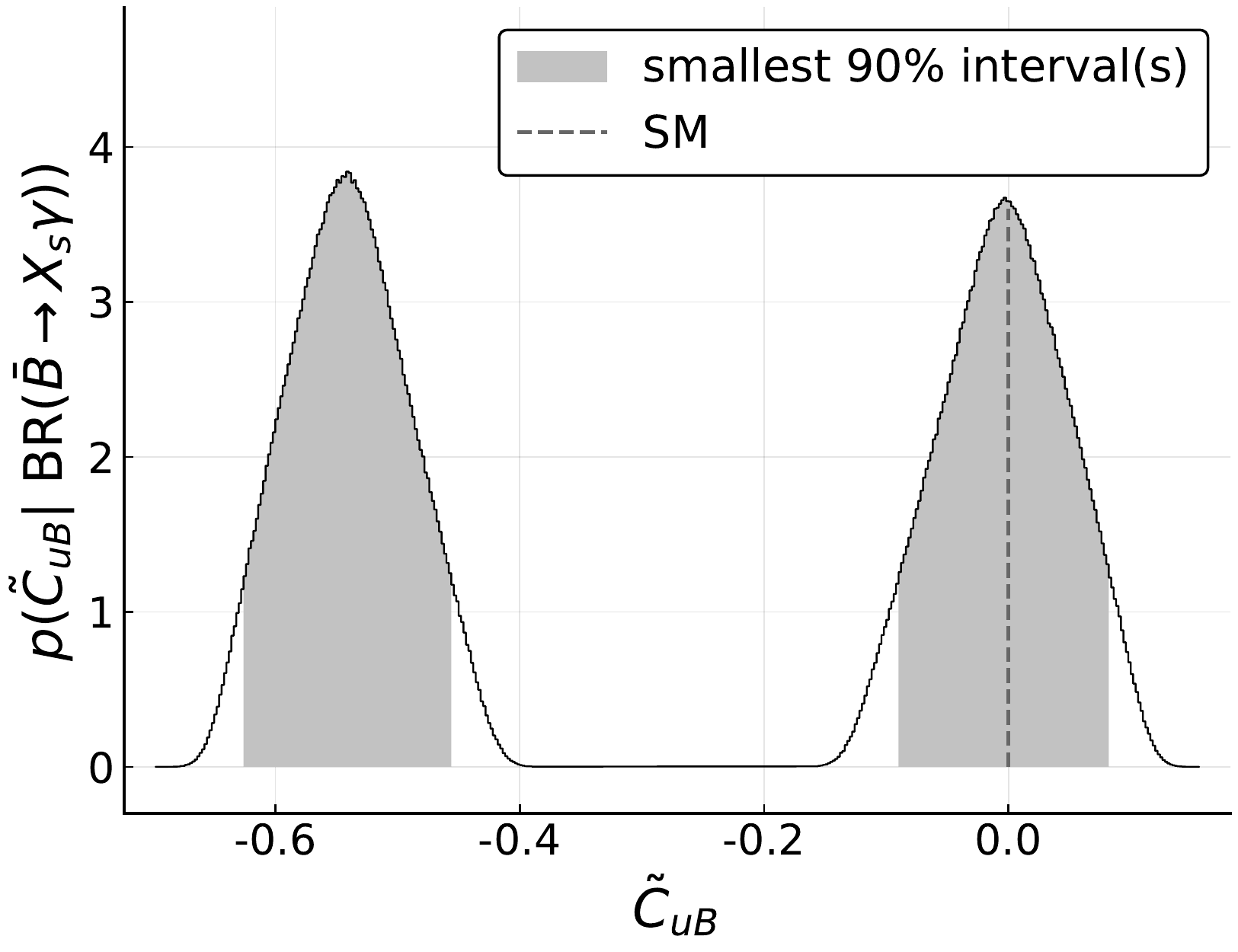}
    \caption{Marginalized posterior probability distribution of $\tilde{C}_{uB}$ from the fit of all three Wilson coefficients to BR($\bar B \rightarrow X_s \gamma $) only. The smallest interval containing \SI{90}{\percent} of the posterior probability and the SM value (dashed line) are indicated}
    \label{Fig:CuB_onlyB}
\end{figure}
The marginalized posterior distribution of $\tilde{C}_{uB}$ is shown in Fig. \ref{Fig:CuB_onlyB}. Two regions for $\tilde{C}_{uB}$ are favored by the fit. 
Comparing with Fig. \ref{Fig:BRSmeftCoeff}, the two regions with the highest probability at about $\tilde{C}_{uB} \approx -0.5$ and $\tilde{C}_{uB} \approx 0.0$ are reasonable since the quadratic shape of BR($\bar B \rightarrow X_s \gamma $) as a function of $\tilde{C}_{uB}$ leads to an agreement with the measurement in these two regions. 
Apparently, without further information, neither of them can be rejected. Indeed, as is well-known, this ambiguity can be resolved by studies of semileptonic $ b \to s \ell^+ \ell^-$ decays \cite{Ali:1999mm}, notably, angular distributions
thereof, whose measurements support the close-to-the-SM branch \cite{Aaij:2013iag}.
Since the purpose of this work is to demonstrate complementarity and feasibility of  a joint bottom and top SMEFT-analysis rather than performing a most global fit, we leave the study of further observables beyond BR($\bar B \rightarrow X_s \gamma $)  and  $\sigma(t\bar t \gamma)$ for future work. 
\par
\subsection{Top physics only \label{sec:t}}
\begin{figure*}[htp]
         \includegraphics[width=0.49\textwidth]{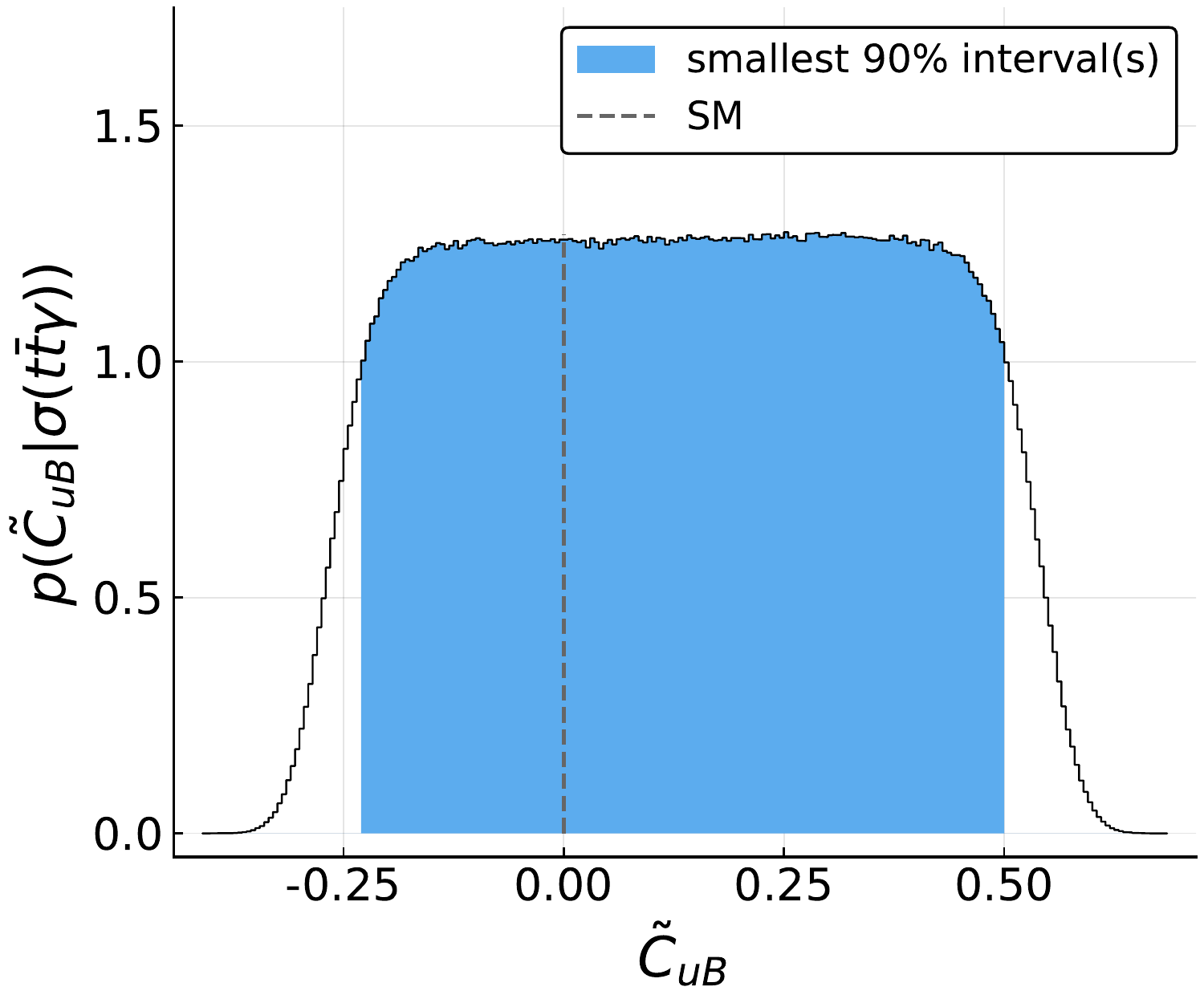}
         \includegraphics[width=0.49\textwidth]{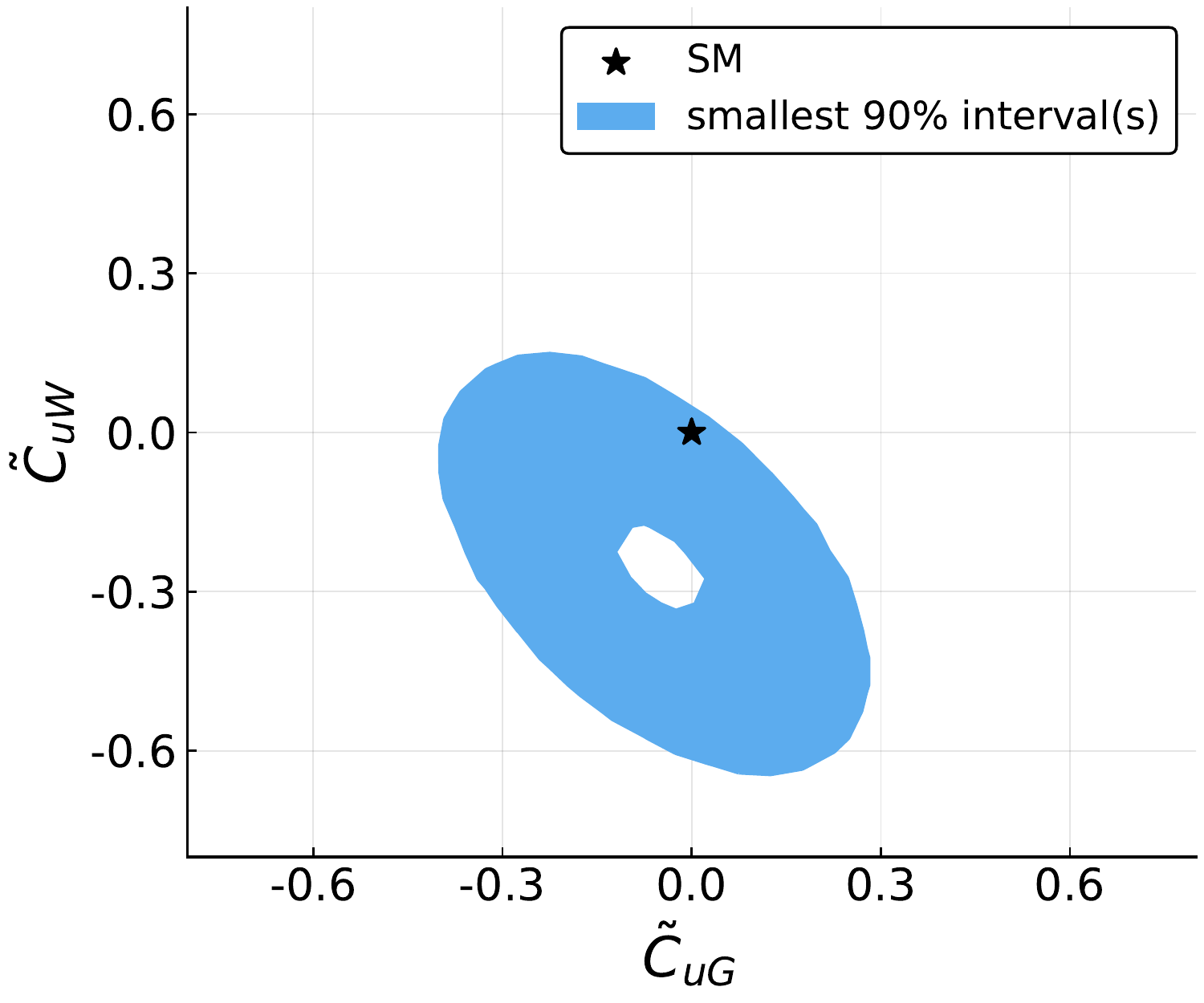}
\caption{Posterior probability distributions for the fit of all three Wilson coefficients using only the measurements of $\sigma(t\bar t \gamma)$. Shown are (left) the marginalized posterior probability distribution of $\tilde{C}_{uB}$ together with the corresponding smallest interval containing \SI{90}{\percent} of the probability and (right) the smallest interval containing \SI{90}{\percent} of the posterior probability for the 2D marginalized distribution of $\tilde{C}_{uG}$ vs. $\tilde{C}_{uW}$. The SM values are indicated}
\label{Fig:results_onlyT}
\end{figure*}

We perform a fit of the Wilson coefficients using $\sigma(t\bar t \gamma)$ only. 
We apply the parametrizations of the single-lepton and dilepton channel fiducial cross sections obtained in Sec. \ref{Sec:CStty} and fit to the corresponding measurements described in Sec. \ref{tta_measurements}. 
Again, all three Wilson coefficients are free parameters of the fit, having uniform prior probability distributions within the range [-1, 1]. 
The resulting marginalized posterior distribution of $\tilde{C}_{uB}$ and the smallest area containing \SI{90}{\percent} of the posterior probability of the 2D marginalized distribution of $\tilde{C}_{uG}$ vs. $\tilde{C}_{uW}$ are shown in Fig. \ref{Fig:results_onlyT}.
With a fit to $\sigma(t\bar t \gamma)$ all three Wilson coefficients can be constrained to a similar extent. The posterior probability distributions of the coefficients have similar shapes and the \SI{90}{\percent} intervals are of comparable size. 
These results are compatible with what is observed in the parabolas shown in Fig. \ref{Fig:tta}. 
When performing the fit considering only the single-lepton or only the dilepton channel measurements as a cross check, very similar results are obtained. This is also expected from Fig. \ref{Fig:tta} as it indicates that both channels have similar  sensitivity to the Wilson coefficients.\par
\subsection{Combined analysis \label{sec:tplusb}} 
\begin{figure*}
    \includegraphics[width=0.49\textwidth]{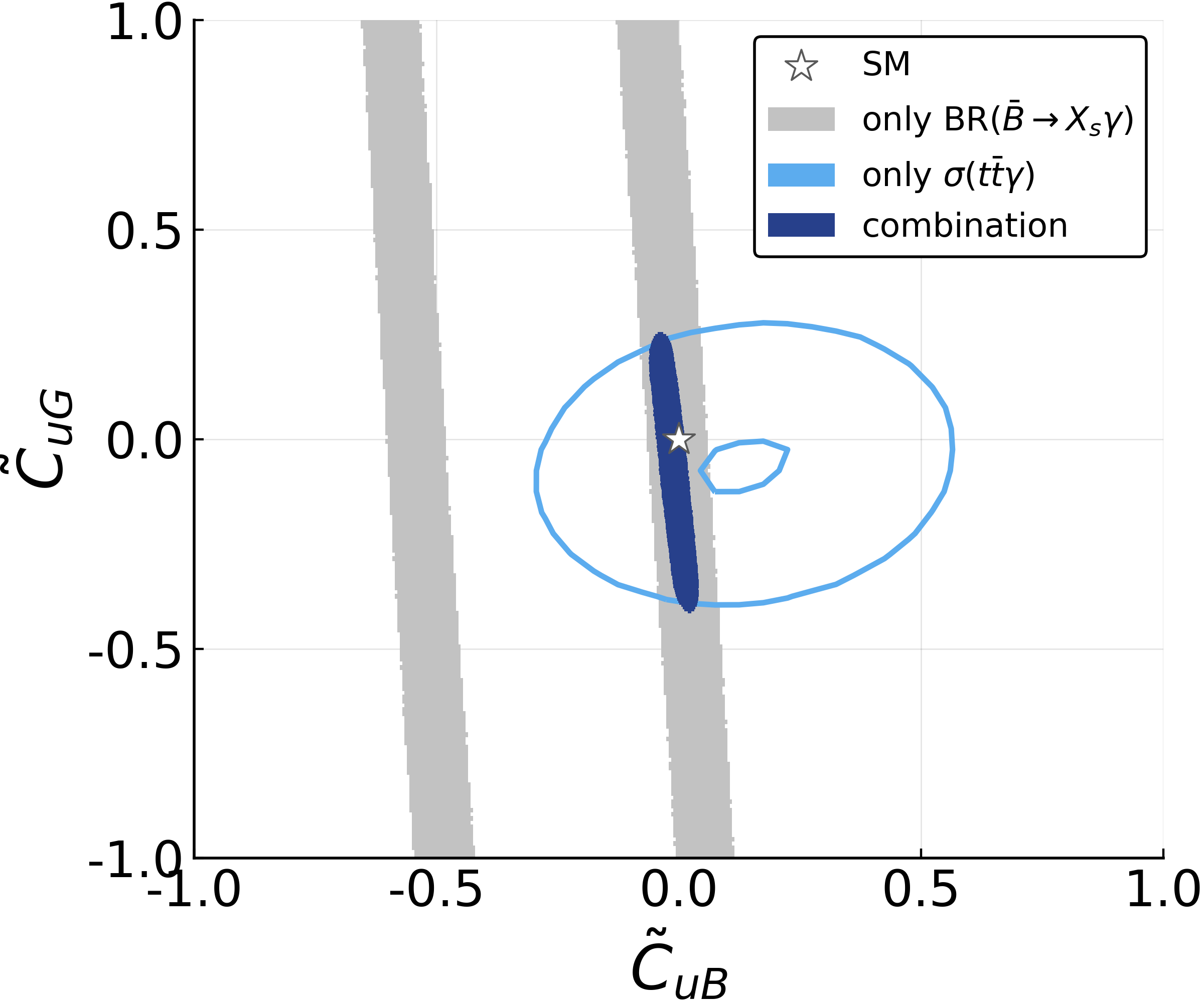}
    \includegraphics[width=0.49\textwidth]{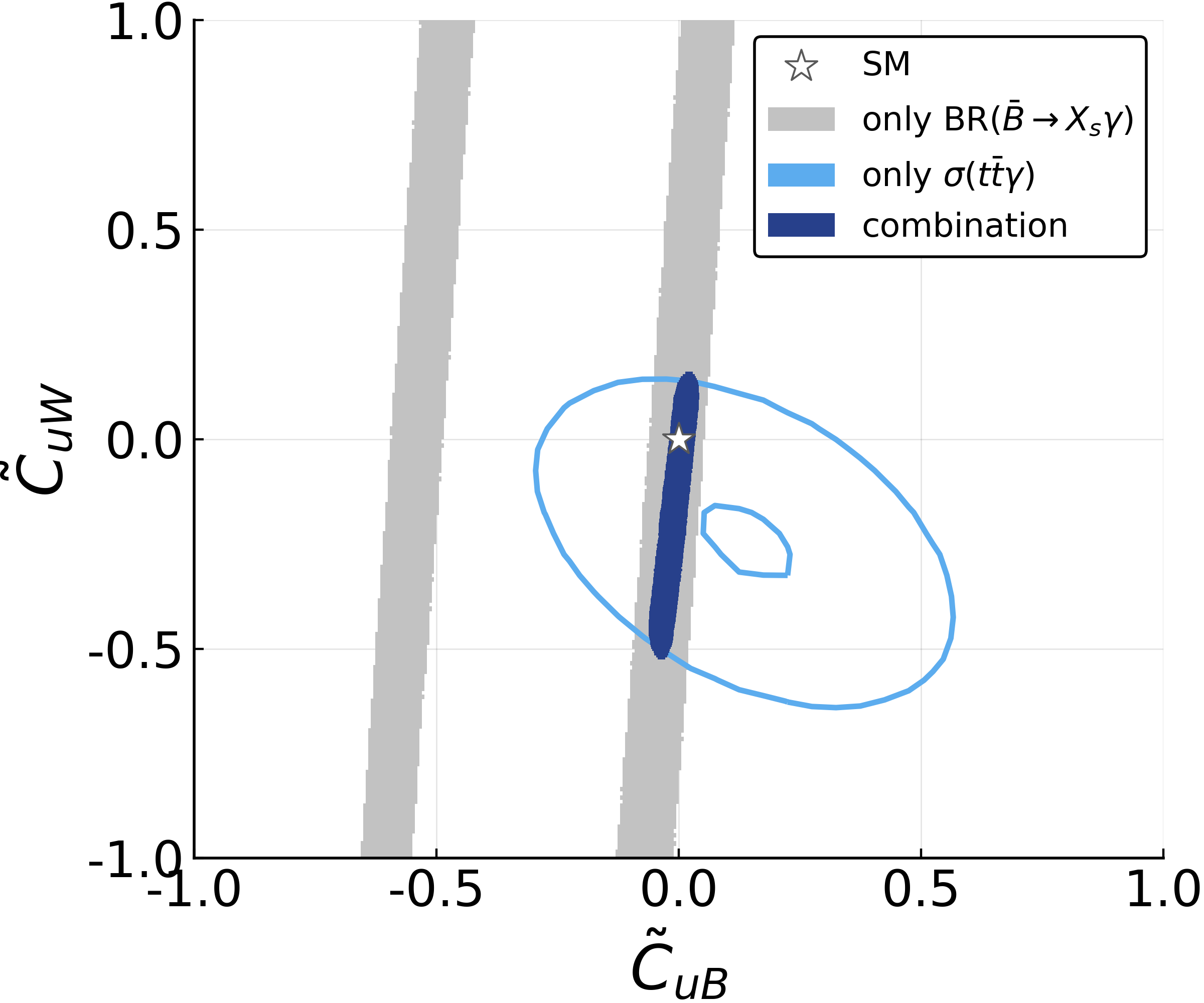}\\[5ex]
    \centering
    \includegraphics[width=0.49\textwidth]{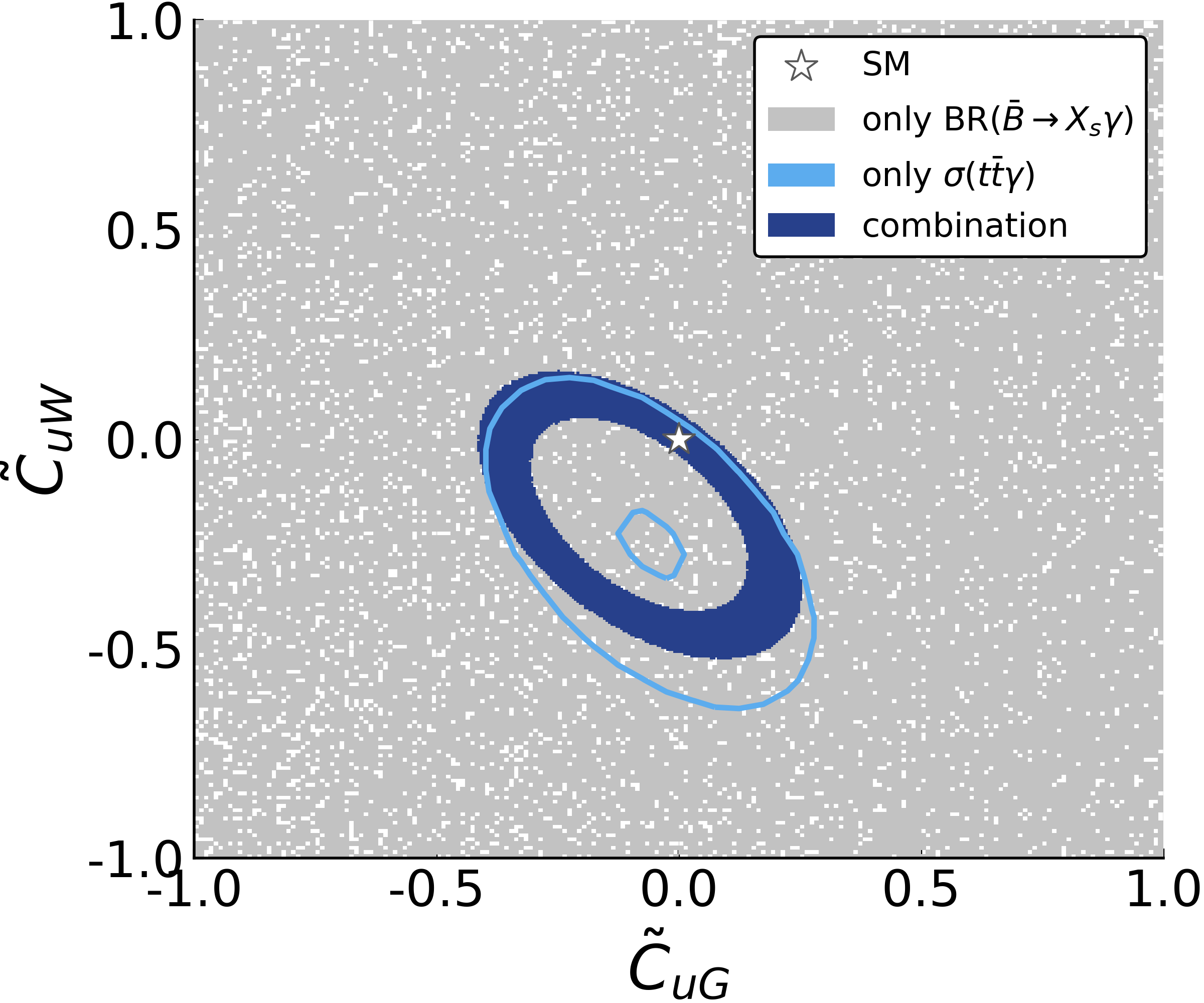}
    \caption{Comparison of the smallest intervals containing \SI{90}{\percent} of the 2D marginalized posterior probability distributions for the fits of all three Wilson coefficients using only the measurement of BR($\bar B \rightarrow X_s \gamma $), only the measurements of $\sigma(t\bar t \gamma)$ and for the combination. The SM values are indicated}
    \label{Fig:comparison2d}
\end{figure*}

For the combined fit, we apply the same uniform priors as in the individual fits and constrain $\tilde{C}_{uB}$, $\tilde{C}_{uG}$ and $\tilde{C}_{uW}$ using both BR($\bar B \rightarrow X_s \gamma $) and $\sigma(t\bar t \gamma)$. 
The resulting smallest areas containing \SI{90}{\percent} of the posterior probability are shown in Fig. \ref{Fig:comparison2d} for the 2D marginalized distributions.
The plots also include the corresponding \SI{90}{\percent} regions from the previously described fits including only one set of observables at a time.
In Fig. \ref{Fig:comparison2d} it is noticeable that the ambiguity in $\tilde{C}_{uB}$, which is observed in the fit including only the BR($\bar B \rightarrow X_s \gamma $) measurement, is resolved in the combined fit. 
It is recognizable that even though the branching fraction measurement alone constrains only $\tilde{C}_{uB}$, in the combination with the  $t\bar t \gamma$ cross sections the constraints on all three Wilson coefficients improve as the sizes of the areas containing \SI{90}{\percent} of the probability decrease in all plots. 
The \SI{90}{\percent} area of the fit using only BR($\bar B \rightarrow X_s \gamma $) in the upper left plot of Fig. \ref{Fig:comparison2d} has a size of \SI{12}{\percent} of the total parameter space $\tilde{C}_{uB} \in [-1, 1]$ and $\tilde{C}_{uG} \in [-1, 1]$ specified by the priors. For the fit considering only $\sigma(t\bar t \gamma)$ the corresponding area is of a similar size, taking up about \SI{11}{\percent} of the allowed space. Due to the orthogonality of the observables, combining top and bottom measurements gives, on the other hand, a \SI{90}{\percent} posterior region reduced by more than an order of magnitude, yielding an area that corresponds to only about \SI{1}{\percent} of the allowed parameter space. The same numbers apply also for the upper right plot of $\tilde C_{uB}$ vs. $\tilde C_{uW}$.
Even in the bottom plot of Fig. \ref{Fig:comparison2d}, which does not directly depend on $\tilde{C}_{uB}$ and is thus not directly constrained by the branching fraction measurement, the \SI{90}{\percent} area is reduced. 
In combination with the BR($\bar B \rightarrow X_s \gamma $) measurement, the  \SI{90}{\percent} area decreases by a factor of 1.9 compared to the fit considering only the $\sigma(t\bar t \gamma)$ measurements. This is a consequence of the reduction of allowed regions in the three-dimensional parameter space.\par
\begin{figure*}[htb]
     \centering
         \includegraphics[width=0.49\textwidth]{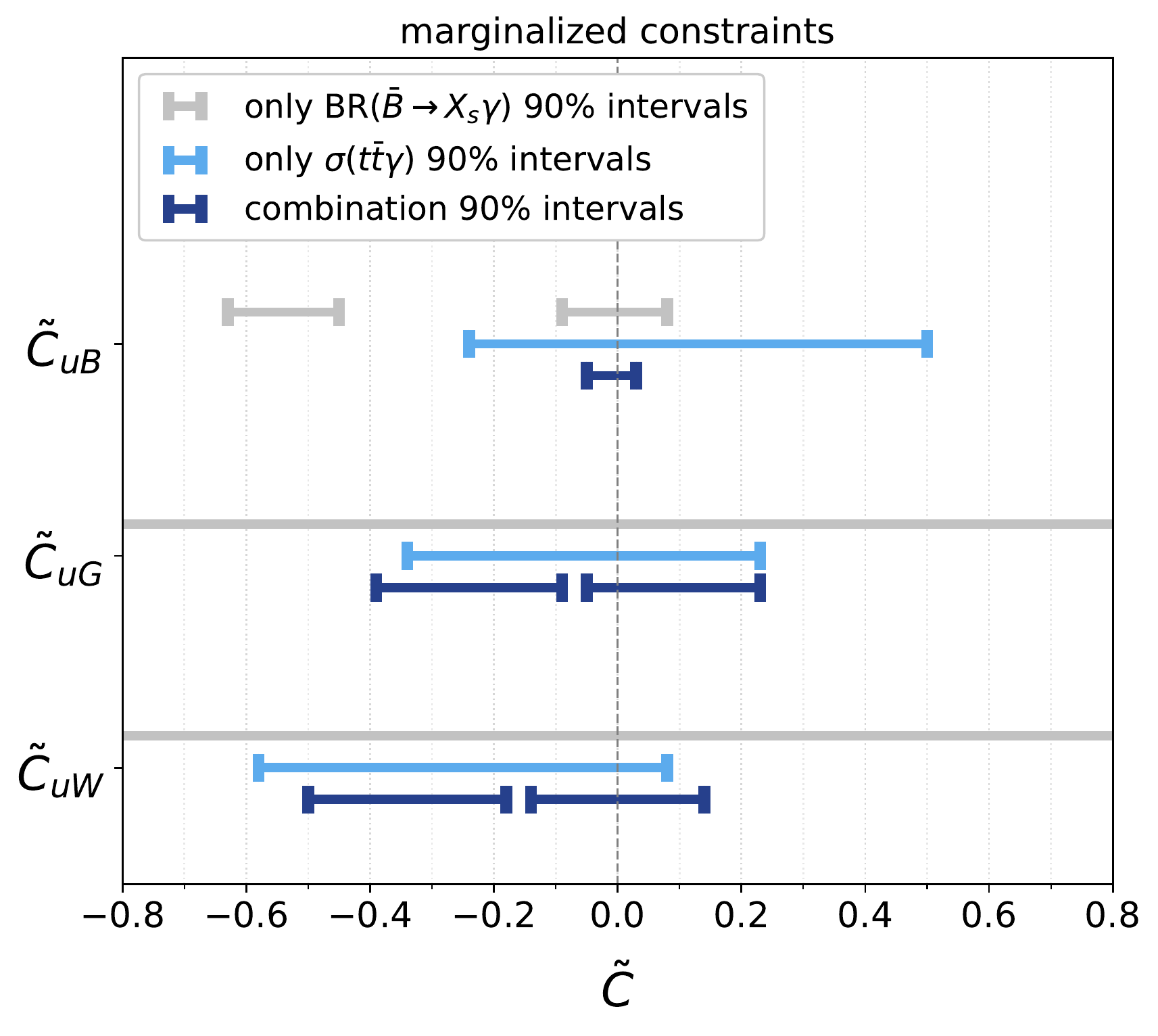}
         \includegraphics[width=0.49\textwidth]{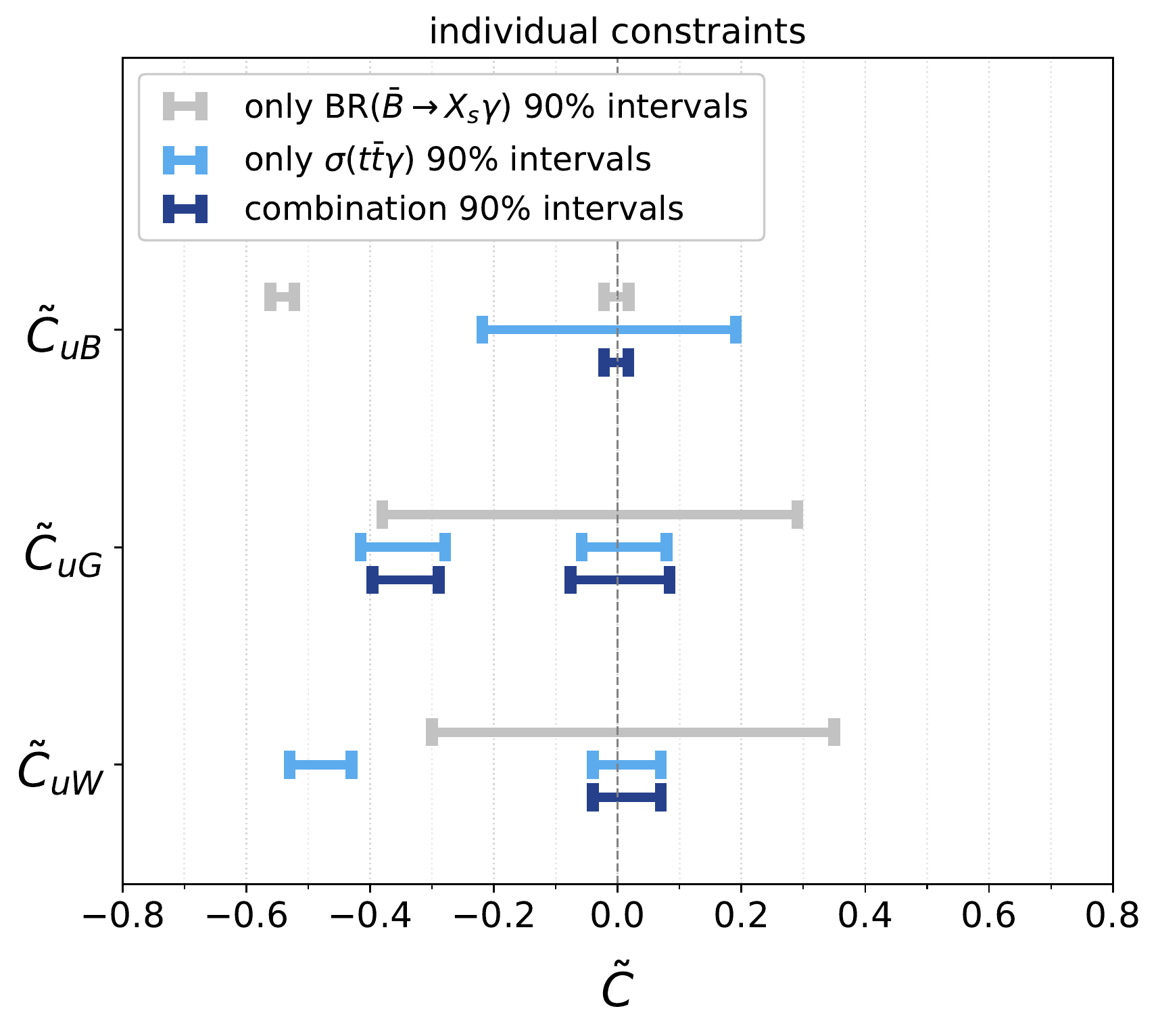}
     \caption{Comparison of the smallest intervals containing \SI{90}{\percent} probability of the 1D marginalized posterior distribution for the fits using only BR($\bar B \rightarrow X_s \gamma $), only $\sigma(t\bar t \gamma)$ and using their combination. Shown are the intervals for (left) the 1D marginalized posterior distributions for the fit of all three Wilson coefficients at a time  and (right) for individual fits of each Wilson coefficient, while the other two coefficients are fixed to zero}
     \label{Fig:comparison}
\end{figure*}

 
A different representation of the same fit results is given in the left plot of Fig. \ref{Fig:comparison}, where the smallest intervals containing \SI{90}{\percent} probability of the 1D marginalized posterior distributions are shown for the combined fit as well as for the fits using only one of the measurements.\par
In the right plot of Fig. \ref{Fig:comparison} the smallest intervals containing \SI{90}{\percent} probability of the 1D marginalized posterior distribution are shown for individual fits in which only one of the Wilson coefficients is allowed to vary at a time, while the other two are fixed to zero.
Overall, a similar behaviour of the results can be observed compared to the fits with three free parameters. 
As there are fewer degrees of freedom in the fits, stronger constraints on the Wilson coefficients can be obtained. 
It is noticeable that in the individual fits not only the ambiguity in $\tilde{C}_{uB}$ can be resolved by the $t\bar t \gamma$ measurement but that also an ambiguity in the top-measurements interval of $\tilde{C}_{uW}$ can be resolved by BR($\bar B \rightarrow X_s \gamma $). \par
As mentioned above, we study the impact of correlations between the systematic and theoretical uncertainties of the single-lepton and dilepton channels of $\sigma(t\bar t \gamma)$.
For this purpose, we perform the combined fit assuming different correlations between the two channels for these uncertainties. 
We vary the correlation coefficient of the systematic uncertainties between values of $-0.9$ and $ 0.9$ as negative correlations are conceivable. The correlation coefficient of the theory uncertainties is varied up to a value of $0.9$ since we do not expect negative correlations for these uncertainties.
When comparing the sizes of the areas containing \SI{90}{\percent} of the marginalized posterior probability to the results assuming uncorrelated uncertainties, we observe only minor changes for the two distributions of $\tilde{C}_{uB}$ vs. $\tilde{C}_{uG}$ and $\tilde{C}_{uB}$ vs. $\tilde{C}_{uW}$. 
We find relative changes in the size of the areas of about \SI{4}{\percent} at maximum and no changes in the general shape or positions compared to the combination shown in the two upper plots of Fig. \ref{Fig:comparison2d}.
As the distribution of $\tilde{C}_{uG}$ vs. $\tilde{C}_{uW}$ is dominantly constrained by the $\sigma(t\bar t \gamma)$ measurements, we observe larger changes due to variations of the correlation coefficients. 
The size of the \SI{90}{\percent} area can change by up to \SI{30}{\percent} for this distribution. Again, the general shape and the positions are not affected but only the width of the ring in the bottom plot of Fig. \ref{Fig:comparison2d} varies.
Therefore, we conclude that even in the presence of correlations between the systematic or theoretical uncertainties of the single-lepton and dilepton channels our previously presented findings are valid.
\par
It should be noted that our focus is to demonstrate how observables from $B$ and top-quark physics can be combined in a single fit of the SMEFT Wilson coefficients. Using only two observables, we do not obtain the most stringent constraints on the coefficients considered. 
Including further observables would certainly improve the constraints. For example, the Wilson coefficients $\tilde C_{uG}$ and $\tilde C_{uW}$ are strongly constrained by the $t\bar t$ production cross section and $W$-boson helicity-fraction measurements, respectively \cite{Buckley:2015lku,Hartland:2019bjb}, whereas measurements of semileptonic $b\rightarrow s \ell^+\ell^-$ decays, especially $B\rightarrow K^{*}\mu^+\mu^-$ angular distributions \cite{Aaij:2013iag}, exclude values $\tilde{C}_{uB}\approx -0.5$ which are allowed by BR($\bar B\rightarrow X_s \gamma$). 
\section{Conclusions} \label{Sec:Conclusion}
Effective theories provide a  systematic toolbox  to exploit multi-observable systems and probe the SM in a model-independent way.
The SMEFT-framework allows to combine data from the precision flavor  and the high energy frontiers.
We exploited synergies between top-quark and $B$-physics measurements from the LHC and precision flavor factories.

Specifically, we performed an exploratory study combining data on the $\bar B \rightarrow X_s \gamma  $ branching ratio and on fiducial $t\bar t \gamma$ production cross sections within SMEFT,
after detailing the ingredients required to connect measurements from different energy scales.
We pointed out that for the processes considered in this work it is necessary to perform a dedicated matching that goes beyond the tree-level matching that is currently 
available in tools.  
Using MC simulations and a particle-level event selection, we performed interpolations of the total $t\bar t \gamma$ production cross section and the fiducial acceptances to parametrize the dependence of the fiducial cross sections on the Wilson coefficients. \par

We demonstrated that due to the different sensitivities of the observables to the SMEFT operators, a combination of the fiducial $t\bar t \gamma$ cross section with the $\bar B \rightarrow X_s \gamma $ branching fraction improves the constraints on the Wilson coefficients (Sec.~\ref{Sec:Fit}). The complementarity of the different observables used in the fit allows to resolve ambiguities and to reduce posterior regions in the marginalized parameter space by up to an order of magnitude.\par
Further, more global analyses of combined top-quark and flavor physics measurements should be pursued in the future with more precise data  expected from 
LHCb \cite{Cerri:2018ypt} and Belle II \cite{Kou:2018nap} and the high-$p_T$-experiments \cite{Atlas:2019qfx}, to decipher physics at higher energies and pursue the  quest for BSM physics.

\begin{acknowledgements}
C.G. is supported by the doctoral scholarship program of the \emph{Studienstiftung des deutschen Volkes}.
\end{acknowledgements}
\appendix
\section{Parameters and experimental input}
\label{App:Parameters}
The parameters used for numerical computations are given in Ref.  \cite{Tanabashi:2018oca}
\begin{align*}
    &m_{t}=(173.1\pm0.4)\,\si{\giga\electronvolt}\,,\\
    &m_t(m_t)=\left(160^{+5}_{-4}\right)\,\si{\giga\electronvolt}\,,\\
    &m_b(m_b)=\left(4.18^{+0.04}_{-0.03}\right)\,\si{\giga\electronvolt}\,,\\
    &m_c(m_c)=\left(1.275^{+0.025}_{-0.035}\right)\,\si{\giga\electronvolt}\,,\\
    &m_s(2\,\si{\giga\electronvolt})=\left(0.095^{+0.009}_{-0.008}\right)\,\si{\giga\electronvolt}\,,\\
    &m_{Z}=91.188\,\si{\giga\electronvolt}\,,\\
    &m_{W}=80.4\,\si{\giga\electronvolt}\,,\\
    &\alpha_s(m_Z)=0.1181\,,\\
    &\alpha_{e}=7.29735257\times10^{-3}\,,\\
    &\sin^2\theta_w(m_Z)=0.2313\,,\\
    &G_F=1.166379\times10^{-5}\,\si{\giga\electronvolt}^{-2}\,.
\end{align*}
The relevant CKM Matrix elements are given in Refs. \cite{Bona:2006ah,UTfit}
\begin{align*}
    V_{tb}&=0.999097\pm0.000024\,,\\
    V_{ts}&=(-0.04156\pm0.00056)\exp[(1.040\pm0.035)\si{\degree}]\,,\\
    V_{cb}&=0.04255\pm0.00069\,.
\end{align*}
The experimental input for the computation of BR($\bar B \rightarrow X_s \gamma $) reads \cite{Alberti:2014yda,Aubert:2004aw}
\begin{align*}
        &C=0.568\pm0.007\pm0.01 \ \,,\\
        &\textmd{BR}(\bar B \rightarrow X_c e\bar\nu)_\textmd{exp}=0.1061\pm0.0017 \ \,.
\end{align*}

\section{Matching condition}
\label{App:Match}
The functions $E_{7}^{uW}$, $F_{7}^{uW}$, $E_{8}^{uW}$ and $F_{8}^{uW}$ are given by
\begin{align*}
    E_{7}^{uW}(x_t)&=
    \frac{-9 x_{t}^3+63 x_{t}^2-61 x_{t}+19}{48 \left(x_{t}-1\right)^3}\\
    &+\frac{\left(3 x_{t}^4-12 x_{t}^3-9 x_{t}^2+20 x_{t}-8\right) \ln \left(x_{t}\right)}{24 \left(x_{t}-1\right)^4}\\
    &+\frac{1}{8}\ln\left(\frac{m_W^2}{\mu_W^2}\right)\,,\\
    F_{7}^{uW}(x_t)&=\frac{x_{t} \left(2-3 x_{t}\right) \ln \left(x_{t}\right)}{4 \left(x_{t}-1\right)^4}
    -\frac{3 x_{t}^3-17 x_{t}^2+4 x_{t}+4}{24 \left(x_{t}-1\right)^3}\,,\\
    E_{7}^{uB}(x_t)&=-\frac{1}{8} \ln \left(\frac{m_W^2}{\mu_W^2}\right)-\frac{\left(x_{t}+1\right)^2}{16 \left(x_{t}-1\right)^2}\\
    &-\frac{x_{t}^2 \left(x_{t}-3\right) \ln \left(x_{t}\right)}{8 \left(x_{t}-1\right)^3}\,,\\
    F_{7}^{uB}(x_t)&=-\frac{1}{8}\,,\\
    E_8^{uW}(x_t)&=\frac{3 x_{t}^2-13 x_{t}+4}{8 \left(x_{t}-1\right)^3}
    +\frac{\left(5 x_{t}-2\right) \ln \left(x_{t}\right)}{4 \left(x_{t}-1\right)^4}\,,\\
    F_8^{uW}(x_t)&=\frac{x_{t}^2-5 x_{t}-2}{8 \left(x_{t}-1\right)^3}+\frac{3 x_{t} \ln \left(x_{t}\right)}{4 \left(x_{t}-1\right)^4}\,,\\
    E_8^{uG}(x_t)&=E_7^{uB}(x_t)\,,\\
    F_8^{uG}(x_t)&=F_7^{uB}(x_t)\,.
\end{align*}

\newpage
\section{Fiducial acceptance of the dilepton channel}\label{Sec:app_acc}
\begin{figure}[htbp]
    \centering
    \includegraphics[width=0.35\textwidth]{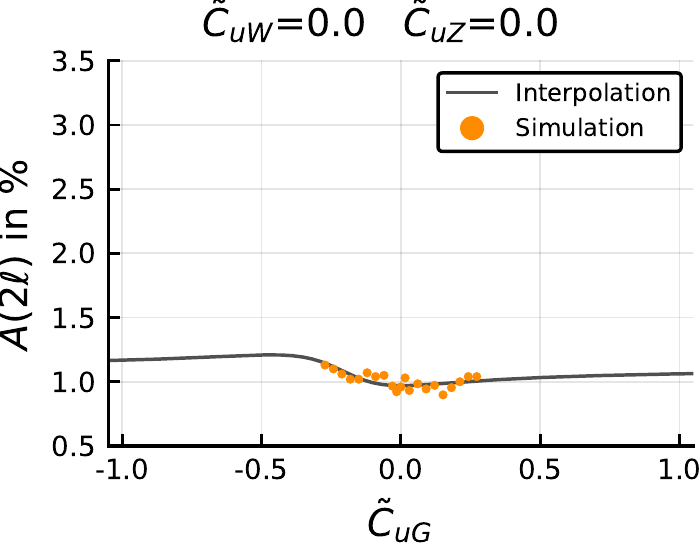}
    \includegraphics[width=0.35\textwidth]{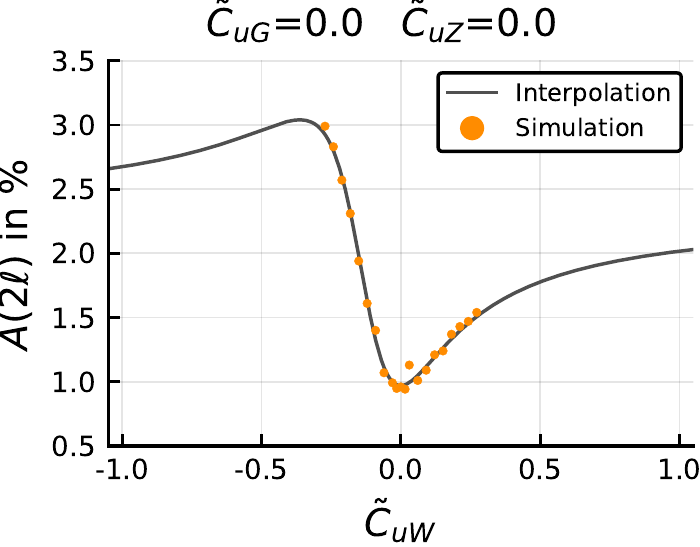}
    \includegraphics[width=0.35\textwidth]{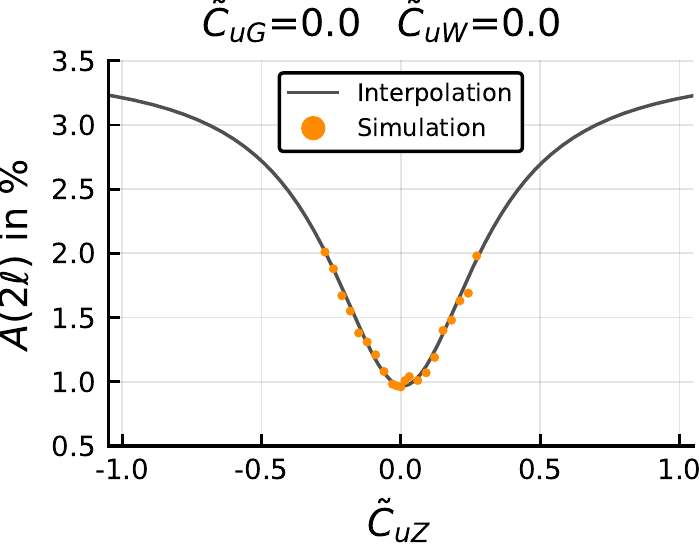}
    \caption{Sampling points and interpolation result for the fiducial acceptance of the dilepton channel $A(2\ell)$, represented as slices of the phase space where only one of the Wilson coefficient is varied at a time, while the others are set to zero.}
    \label{Fig:acceptance_2l}
\end{figure}
\FloatBarrier

\end{document}